\documentclass[%
reprint, 
superscriptaddress,
amsmath,amssymb,
aps,
prx,
]{revtex4-2}
\usepackage{graphicx}
\usepackage{epstopdf}
\usepackage{bm}
\usepackage{hyperref}
\usepackage{comment}
\usepackage{color}
\usepackage{physics}
\usepackage{amsmath}
\usepackage{cancel}
\usepackage{tikz}
\usepackage{enumitem}
\usepackage{wrapfig}
\usepackage{bbold}
\usepackage{dsfont}
\usepackage[normalem]{ulem}
\usepackage{subcaption}

\hypersetup{%
   pdfpagemode=UseNone, 
   pdfstartpage=1,
   pdfmenubar=true,
   pdftoolbar=true,
   colorlinks = true,
   linkcolor=blue,
   citecolor=blue,
   urlcolor=blue,
   bookmarksopen=false}

\definecolor{fuchsia}{rgb}{1.0, 0.0, 1.0}
\definecolor{maroon}{rgb}{0.788, 0.0, 0.086}
\definecolor{ao}{rgb}{0.0, 0.5, 0.0}

\begin{document}

\title{Charge and spin orders in the $t$-$U$-$V$-$J$ model: a slave-spin-1 approach}

\author{Olivier~Simard}
\affiliation{Collège de France, 11 place Marcelin Berthelot, 75005 Paris, France}
\affiliation{CPHT, CNRS, École Polytechnique, IP Paris, F-91128 Palaiseau, France}
\author{Michel~Ferrero}
\affiliation{Collège de France, 11 place Marcelin Berthelot, 75005 Paris, France}
\affiliation{CPHT, CNRS, École Polytechnique, IP Paris, F-91128 Palaiseau, France}
\author{Thomas~Ayral}
\affiliation{CPHT, CNRS, École Polytechnique, IP Paris, F-91128 Palaiseau, France}
\affiliation{Eviden Quantum Laboratory, 78300 Les Clayes-sous-Bois, France}

\date{\today}

\begin{abstract}
Strongly-correlated fermion systems on a lattice have been a subject of intense focus in the field of condensed-matter physics. These systems are notoriously difficult to solve, even with state-of-the-art numerical methods, especially in regimes of parameters where degrees of freedom compete or cooperate at similar energy and length scales. Here, we introduce a spin-1 slave-particle technique to approximately treat the $t$-$U$-$V$-$J$ fermionic model at arbitrary electron dopings in an economical manner. This formalism respectively maps the original charge and spin degrees of freedom into effective pseudo-spin and pseudo-fermion sectors, which are treated using a self-consistent cluster mean-field method. We study the phase diagram of the model under various conditions and report the appearance of charge and spin stripes within this formalism. These stripes are a consequence of the cluster mean-field treatment of the pseudo-particle sectors and have not been detected in previous slave-particle studies. The results obtained agree qualitatively well with what more reliable numerical methods capture.
\end{abstract}

\pacs{}

\maketitle


\section{Introduction}
\label{sec:intro}

Many-body quantum systems remain remarkably challenging to simulate, particularly in regimes where several degrees of freedom compete---or cooperate---at comparable energy and length scales. Furthermore, across temperature and doping, the hierarchy of dominant fluctuations can vary dramatically: for example, in the square lattice Hubbard model, close to the pseudogap regime, spin-density-wave (SDW), charge-density-wave (CDW), and $d$-wave superconductivity (dSC) channels all compete~\cite{PhysRevB.100.094506,doi:10.1126/science.ade9194,PhysRevLett.88.117001,PhysRevLett.113.046402}, giving rise to a rich landscape of intertwined orders~\cite{Keimer2015}. Some experimental works indeed measure SDW/CDW instabilities close to dSC at low temperatures in specific cuprate compounds~\cite{PhysRevB.54.7489,Tranquada1995,PhysRevLett.85.1738}, which can be modeled by the square-lattice Hubbard model. For example, small changes in the electronic band structure can favor or not dSC over CDWs~\cite{PhysRevX.10.031016,doi:10.1126/science.adh7691,roth2025superconductivitytwodimensionalhubbardmodel}.

To understand stripe formation and other emergent phenomena in doped Mott insulators~\footnote{In this work, by doped Mott insulator, we mean a bad metal or a bad insulator.}, a broad range of theoretical and numerical approaches has been developed. Among the most reliable are nonperturbative many-body techniques, including the density-matrix renormalization group (DMRG)~\cite{RevModPhys.77.259,PhysRevLett.69.2863}, finite-size projected entangled pair states (PEPS)~\cite{verstraete2004renormalizationalgorithmsquantummanybody,PhysRevA.81.052338}, variational quantum Monte Carlo (VQMC)~\cite{PhysRev.138.A442,PhysRevLett.60.1719,doi:10.1126/science.aag2302}, auxiliary-field quantum Monte Carlo (AFQMC)~\cite{PhysRevLett.74.3652,PhysRevLett.90.136401}, and dynamical mean field theory (DMFT)~\cite{PhysRevLett.62.324,PhysRevB.45.6479,RevModPhys.68.13}. These methods have provided crucial insights into intertwined charge and spin textures in correlated lattice fermion systems~\cite{doi:10.1126/science.adh7691,doi:10.1126/science.aam7127,sharma2025comparingsymmetrizeddeterminantneural,doi:10.1073/pnas.2122059119,PhysRevB.97.075112,PhysRevB.89.155134,mushkaev2025spinstripeshubbardmodelcombined,PhysRevLett.80.1272,PhysRevLett.91.136403,PhysRevX.13.011007,Abram_2017,PhysRevB.98.205132,PhysRevB.58.13506,r4q9-4yvj}. However, their applicability to study symmetry-broken phases is often limited by the computational cost when the system size grows.

This limitation has partly motivated the development of simpler, yet physically intuitive, approximate frameworks capable of accessing larger system sizes and longer length scales. Cluster mean-field theories have been employed to expose the low-energy structure responsible for symmetry-broken phases~\cite{PhysRevB.65.041102,PhysRevB.108.035139}, however they neglect the effects of correlations. To selectively include correlations in a physical subspace, many slave-particle constructions exist~\cite{PhysRevB.29.3035,PhysRevB.70.035114,PhysRevB.81.035106,PhysRevB.81.155118,PhysRevLett.57.1362,PhysRevB.101.245437,PhysRevB.56.12909}. By explicitly separating charge and spin degrees of freedom, and mapping one of these physical subspaces into an effective (noninteracting) cluster mean-field problem, while the other is dealt with methods that capture correlations in the bulk, these cluster mean-field slave-particle approaches offer a conceptually appealing route to rationalize the emergence of long-range inhomogeneous states. Nevertheless, they must be critically benchmarked against more accurate methods to assess which physical ingredients they faithfully capture.

Building on this context, the goal of the present work is to assess what slave-spin theory predicts for stripe physics and for the emergence of doped Mott-insulating behavior. To this end, we study the $t$-$U$-$V$-$J$ model~\cite{Abram_2017} on the square lattice using a slave-spin-$1$ construction. The pseudo-fermion (spin) sector is treated using inhomogeneous mean-field theory, while the pseudo-spin-$1$ (charge) sector is solved via DMRG, ensuring an unbiased treatment of charge correlations. Because our focus is on the doping dependence of CDW/SDW textures and on the metal--insulator transition encoded in the charge sector, the $\mathbb{Z}_2$ slave-spin formalism~\cite{PhysRevB.81.155118} is insufficient, and we therefore employ the enlarged pseudo-spin-$1$ local Hilbert space.

Our approach allows us to spatially resolve charge and spin textures, study stripe formation across a broad doping range, and identify how the underlying interactions---on-site repulsion $U$, nearest-neighbor (NN) Coulomb repulsion $V$, NN spin exchange $J$, and hole doping $\delta$---control the onset of stripe order and insulating behavior. The auxiliary coupling $J$ is introduced to seed antiferromagnetic Néel order in the pseudo-fermion sector. Because the slave-spin representation maps the charge sector to an effective spin model, our results also delineate which aspects of stripe physics could, in principle, be realized or benchmarked directly on Rydberg-atom arrays~\cite{PhysRevB.109.174409,2025hybridquantumclassicalanalogsimulation}.

The rest of the paper is organized as follows. Section~\ref{sec:Methods} introduces the model and the slave-spin-$1$ framework. 
Results on the stripe formation and the metal-insulator transition are given in Sections~\ref{sec:self_consistent_mf_solution}.
Finally, in Sec.~\ref{sec:exact_solution}, we compare the slave-spin results with direct DMRG simulations. 

\section{Methods}
\label{sec:Methods}

To be able to map a fermion many-body system into an interacting pseudo-spin system, we split up the most relevant degrees of freedom in a self-consistent manner, effectively decoupling the charge and spin channels. We denote the sector that treats the charge (spin) degrees of freedom the pseudo-spin (pseudo-fermion) sector. This is motivated by the observation that the charge and spin responses are often quite different in the low-energy description of strongly interacting systems; for example, close to the Mott transition, charge correlations dominate over the spin ones and their degrees of freedom get \textit{frozen}. 

To achieve the latter, we resort to slave-particle mean-field theory which leads, on the one hand, to a quartic quantum pseudo-spin-1 model that captures the strongly-correlated physics in the pseudo-spin sector, while, on the other hand, leading to a description of the physical spin degrees of freedom as renormalized spin-1/2 pseudo-fermions.


\subsection{$t$-$U$-$V$-$J$ model}
\label{sec:tJUV_model}

We consider the single-orbital Hamiltonian featuring spin exchange coupling $J$, on-site Coulomb interaction $U$ and NN Coulomb interaction $V$:

\begin{align}
\label{eq:tUVJ_Ham}
\hat{H}_{tUVJ} &= -t\sum_{\langle ij\rangle,\sigma}\left(\hat{c}^{\dagger}_{i,\sigma}\hat{c}_{j,\sigma} + \text{H.c}\right) + \frac{U}{2}\sum_{i,\sigma}\hat{n}_{i,\sigma}\hat{n}_{i,-\sigma}\notag\\
&+ J\sum_{\langle ij\rangle} \ \hat{\mathbf{S}}^c_i\cdot\hat{\mathbf{S}}^c_j + V\sum_{\langle ij\rangle,\sigma\sigma^{\prime}}\hat{n}_{i,\sigma}\hat{n}_{j,\sigma^{\prime}},
\end{align}
where $\hat{c}^{(\dagger)}_{\sigma}$ annihilates (creates) a fermion with spin projected quantum number $\sigma\in \{\uparrow,\downarrow\}$, $\hat{n}_{i,\sigma}=\hat{c}^{\dagger}_{i,\sigma}\hat{c}_{i,\sigma}$, and the SU(2) spin operator $\hat{\mathbf{S}}^c_i = \sum_{\sigma, \sigma'}\hat{c}^{\dagger}_{i,\sigma}\boldsymbol{\tau}_{\sigma\sigma'}\hat{c}_{i,\sigma'}$, and $\boldsymbol{\tau}$ the vector of Pauli spin matrices: $\boldsymbol{\tau}\equiv \left(\sigma_x,\sigma_y,\sigma_z\right)$. The electron hopping is limited to NN hoppings, denoted $t$, and sets the energy scale of the system. $J$ represents the NN spin exchange coupling. The NN Coulomb repulsion $V$ is the same regardless of the spin species. The notation $\langle\cdot\rangle$ signifies that the sum runs over the set of unique NN pairs of lattice sites. In Eq.~\eqref{eq:tUVJ_Ham}, in the case where $V=J=0$, one recovers the Hubbard model.

We introduce a $J$ term explicitly to be able to study stripes within the slave-spin method. Indeed, the slave-spin-1 method we are about to use does not capture the superexchange coupling $J\sim \frac{t^2}{U}$ induced by Hubbard interactions.
This is related to the fact that pseudo-spin degrees of freedom resolve only the different charge states locally, as opposed to the spin states. We therefore explicitly include a $J>0$ term in the Hamiltonian to impose a tendency towards antiferromagnetism (AFM) and thus, upon doping, allow for the formation of stripes. As will be detailed later, the $J$ term of the $t$-$U$-$V$-$J$ model passes the 2-body spatial correlations across the pseudo-spin and pseudo-fermion sectors. This back-and-forth exchange of spatially-resolved 2-body correlators through $J$ turns out to be very important to observe stripes. The NN Coulomb interaction $V$, although not necessary to obtain stripes, is also considered in a dedicated section (Sec.~\ref{sec:V_effect}); it does promote further the establishment of stripes. 


\subsection{Slave-particle formalism}
\label{sec:slave_rotor_formalism}

A great wealth of slave-particle formalisms exists, making use of more or fewer auxiliary degrees of freedom to decouple the relevant physical degrees of freedom~\cite{PhysRevB.81.035106,PhysRevB.81.155118,PhysRevLett.57.1362,PhysRevB.29.3035}.
Here, we use a slave-spin-1 method, which is a relatively economical slave-particle theory.

\subsubsection{Slave-spin mapping}

The slave-spin-1 method consists in mapping the original fermionic operators $\hat{c}^{(\dagger)}$ to pseudo-fermion $\hat{f}^{(\dagger)}$ and pseudo-spin-1 ladder operators $\hat{S}^{\pm}$:

\begin{align}
\label{eq:operator_mapping}
\begin{cases}
\hat{c}^{\dagger}_{i,\sigma} = \hat{f}^{\dagger}_{i,\sigma}\hat{S}^{-}_i
\\
\hat{c}_{i,\sigma} = \hat{f}_{i,\sigma}\hat{S}^{+}_i.
\end{cases}
\end{align}
%
States of the Hilbert space are thus split up into a pseudo-fermion part $\ket{\cdot}_f$ and a pseudo-spin part $\ket{\cdot}_{s}$.
Because of the enlarged size of the Hilbert space, we need constraints to select out the physical states among the states in the enlarged Hilbert space. We choose the following constraints on the operators in Eq.~\eqref{eq:operator_mapping}:
\begin{align}
\label{eq:constraint_eqs_2}
\begin{cases}
\hat{S}_i^{z} \equiv \hat{1} - \sum_{\sigma}\hat{n}^f_{i,\sigma}
\\
\sum_{\sigma}\hat{n}_{i,\sigma} \equiv \sum_{\sigma}\hat{n}^f_{i,\sigma}.
\end{cases}
\end{align}
This allows us to define $\langle\hat{S}_i^{z}\rangle_s\equiv\delta$ as the doping in holes (electrons) when positive (negative); clearly, from the first equation of Eq.~\eqref{eq:constraint_eqs_2}, $\delta = \langle \hat{S}^z_i\rangle_s < 0$ is related to electron doping, while a positive value implies hole doping. Half-filling corresponds to $\delta = 0$. The notation $\langle\cdot\rangle_{s(f)}$ means that the expectation value is taken in the pseudo-spin (pseudo-fermion) sector.

These constraints are obeyed by the so-called physical states:
\begin{align}
\label{eq:original_fermion_mappings}
\begin{cases}
\ket{0} \to \ket{0}_f\ket{1}_s\\
\ket{\sigma} \to \ket{\sigma}_f\ket{0}_s\\
\ket{\uparrow\downarrow} \to \ket{\uparrow\downarrow}_f\ket{-1}_s.
\end{cases}
\end{align}
In this representation, the $t$-$U$-$V$-$J$ Hamiltonian \eqref{eq:tUVJ_Ham} reads, using the prescriptions \eqref{eq:operator_mapping},
\begin{align}
\label{eq:t_U_J_model_slave_representation}
\hat{H}_{tUVJ} &= -t\sum_{\langle ij\rangle,\sigma}\left(\hat{f}_{i,\sigma}^{\dagger}\hat{f}_{j,\sigma}\hat{S}^{-}_i\hat{S}^{+}_j + \text{H.c}\right)\notag\\
&+ \frac{U}{2}\sum_{i}\hat{S}^z_i(\hat{S}^z_i-1) + J\sum_{\langle ij\rangle} \ \hat{\mathbf{S}}^f_i\cdot\hat{\mathbf{S}}^f_j\hat{S}^-_i\hat{S}^+_i\hat{S}^-_j\hat{S}^+_j\notag\\
&+ V\sum_{\langle ij\rangle} \left(\hat{S}^z_i-1\right)\left(\hat{S}^z_j-1\right).
\end{align}

All the algebraic manipulations done hitherto are exact and Eq.~\eqref{eq:t_U_J_model_slave_representation} is solely a recasting of Eq.~\eqref{eq:tUVJ_Ham} when the constraints are satisfied.

Slave-spin-1 theory is more economical than other slave-particle mean-field techniques, since $3$ local slave-spin degrees of freedom are required for a SU$(2)$ spin-symmetric Hamiltonian. This is similar to slave-rotors~\cite{PhysRevB.76.195101,PhysRevB.66.165111,PhysRevB.70.035114} (when truncating the number of excitations of the rotors to $3$), and more economical than slave-boson techniques that require $4$ additional bosonic degrees of freedom~\cite{PhysRevLett.57.1362}.


\subsubsection{Mean-field decoupling of charge and spin}
\label{sec:mean_field_decoupling_charge_spin}

Hamiltonian \eqref{eq:t_U_J_model_slave_representation} is daunting to solve, therefore we decouple the pseudo-spin and pseudo-fermion sectors. To proceed with the `separation' of the pseudo-spin and pseudo-fermion subspaces, we assume that the total many-body wave function of the interacting system can be factorized $\ket{\Psi} = \ket{\Psi_f}\ket{\Psi_s}$.
Such a factorized form leads to a simplified Hamiltonian expression $\hat{H}_{tUVJ} \approx \hat{H}^{s}_{tUVJ} + \hat{H}^{f}_{tUVJ}$, with:
\begin{subequations}
  \begin{align}
    \begin{aligned}
    \label{eq:decoupled_equations_Ham_theta}
      \hat{H}^{s}_{tUVJ} &= -t\sum_{\langle ij \rangle}\underbrace{\sum_{\sigma}\langle \hat{f}_{i,\sigma}^{\dagger}\hat{f}_{j,\sigma}\rangle_{f}}_{\chi_{ij}}\left(\hat{S}^{-}_i\hat{S}^{+}_j+\text{H.c}\right)\\ 
      &+ \frac{U}{2}\sum_i\hat{S}^{z}_{i}\left(\hat{S}^{z}_{i}-1\right)\\ 
      &+ J\sum_{\langle ij\rangle} \ \langle\hat{\mathbf{S}}^f_i\cdot\hat{\mathbf{S}}^f_j\rangle_f \ \hat{S}_i^-\hat{S}^+_i\hat{S}_j^-\hat{S}^+_j \\
      &+ V\sum_{\langle ij\rangle}\left(\hat{S}^{z}_{i}-1\right)\left(\hat{S}^{z}_{j}-1\right)- \mu_s\sum_{i}\hat{S}^z_{i}\phantom{====},
    \end{aligned}\\
    \begin{aligned}
    \label{eq:decoupled_equations_Ham_f} 
      \hat{H}^f_{tUVJ} &= -t\sum_{\langle ij \rangle,\sigma}\underbrace{\langle \hat{S}^{-}_i\hat{S}^{+}_j\rangle_{s}}_{B_{ij}}\left(\hat{f}_{i,\sigma}^{\dagger}\hat{f}_{j,\sigma} + \text{H.c}\right)\\
      &+ J\sum_{\langle ij\rangle} \ \hat{\mathbf{S}}^f_i\cdot\hat{\mathbf{S}}^f_j \ \langle\hat{S}_i^-\hat{S}^+_i\hat{S}_j^-\hat{S}^+_j\rangle_s - \mu_f\sum_{i,\sigma}\hat{n}^f_{i,\sigma}.
    \end{aligned}
  \end{align}
\end{subequations}
where $\mu_f$ and $\mu_{s}$ are chemical potentials introduced to enforce the constraints laid out in Eq.~\eqref{eq:constraint_eqs_2}. In Eqs.~\eqref{eq:decoupled_equations_Ham_f} and \eqref{eq:decoupled_equations_Ham_theta}, we defined the 2-point correlation functions $B_{ij}$ and $\chi_{ij}$, respectively.

These two Hamiltonians contain only pseudo-spin and pseudo-fermion degrees of freedom. The pseudo-spin Hamiltonian is a variant of the XXZ Hamiltonian. It is still a complicated many-body problem, but without the complications arising from fermions.
It features a nearest-neighbor, in-plane coupling that favors (when $-t \chi_{ij}\leq 0$) an in-plane ordering of the pseudo-spins; an out-of-plane field $U/2$ that penalizes states $S^z_i =-1$ (namely double occupancies) and competes with the in-plane ordering tendency of the first term; a $J$ term that, when the pseudo-fermions have AFM tendencies ($\langle\hat{\mathbf{S}}^f_i\cdot\hat{\mathbf{S}}^f_j\rangle_f \leq 0$), penalize bonds with both $S^z_i = 1$ and $S^z_j = 1$ (namely holes); finally, the $V$ term penalizes bonds with NN occupancies. 
We are going to solve this model using an inhomogeneous cluster mean-field method, where the cluster ground state is obtained via DMRG, with clusters of size up to $4 \times 32$.
This approach goes further than most previous studies, which usually resort to a single-site mean-field treatment of the model (clusters of size $1$), and therefore completely neglect spatial correlations.

As for the pseudo-fermion Hamiltonian: except for the $J$ term, it is quadratic in the pseudo-fermion fields. We are going to approximate the $J$ term at the mean-field level, making this Hamiltonian quadratic and therefore solvable in polynomial time.

We note that the mean-field decoupling implies that the constraints are satisfied on average in each sector (see Appendix~\ref{sec:constraint_validity}). Note that the 4-point terms appearing in both the pseudo-fermion sector Eq.~\eqref{eq:decoupled_equations_Ham_f} and pseudo-spin sector Eq.~\eqref{eq:decoupled_equations_Ham_theta} conserve the particle number operator (the 4-point term in Eq.~\eqref{eq:decoupled_equations_Ham_theta} commutes with $\hat{S}^z$).

\begin{figure}[h!]
  \centering
    \includegraphics[scale=0.7]{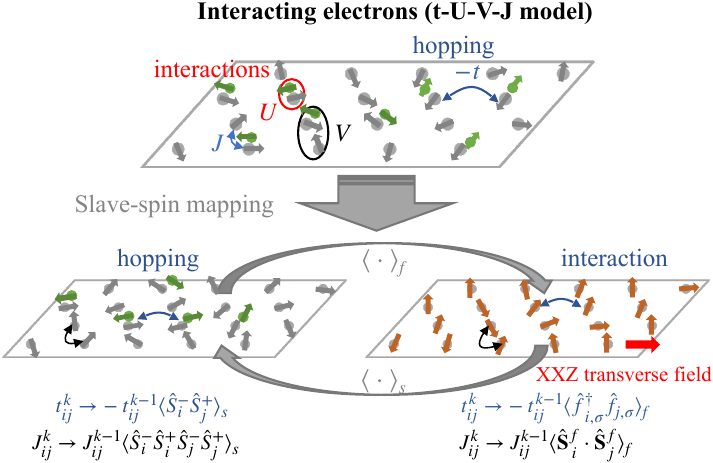}
      \caption{\textbf{Flow chart of the slave-particle spin-1 scheme} to decouple the charge and spin degrees of freedom self-consistently. $t^k_{ij}$ ($J^k_{ij}$) represents the electron (spin) NN tunneling matrix at iteration $k$, renormalized appropriately in each sector. The matrices $t^k_{ij}$ and $J^k_{ij}$ inherit their spatial dependence from the correlation functions that renormalize the bare parameters $t$ and $J$. $U$ is the on-site interaction and $V$ the NN Coulomb interaction. The various correlation functions renormalize the microscopic couplings.}
  \label{fig:flow_chart_scheme}
\end{figure} 

In Fig.~\ref{fig:flow_chart_scheme}, we summarize our self-consistent slave-particle mean-field procedure. 
The spin-charge decoupling breaks down the $t$-$U$-$V$-$J$ model into a quantum XXZ spin-1 model with a transverse field to treat the charge degrees of freedom (pseudo-spins), and a free renormalized pseudo-fermion problem representing the spin degrees of freedom.
The 2-point and 4-point correlation functions
are evaluated in their respective sectors to update their respective counterpart, thereby renormalizing the microscopic couplings self-consistently (see the arrows looping back and forth of Fig.~\ref{fig:flow_chart_scheme}).

In Appendix~\ref{sec:constraint_validity}, we show the validity of the \textit{Lagrange multipliers} $\mu_s$ and $\mu_f$ by illustrating the Coulomb staircase.
Note that in slave-particle mean-field theories, by breaking down the original physical fermionic field operators into a pseudo spin and a pseudo fermion, one introduces a local gauge-field degree of freedom, whose fluctuations---that would normally lead to the confinement of the composite pseudo-particles---are neglected in this work.

\subsection{Self-consistent cluster mean-field approximation}
\label{sec:self_consistent_cluster_approximation_main}

In this section, we explain how we solve for the ground states of $\hat{H}^s_{tUVJ}$ and $\hat{H}^f_{tUVJ}$. We resort to a self-consistent cluster mean-field approach, which is inspired by the methods used in Refs.~\cite{PhysRevB.76.195101,PhysRevB.81.035106}. For the purpose of completeness and clarity, we schematically introduce the self-consistent cluster mean-field approach using the slave-spin-1 formalism covered in Sec.~\ref{sec:slave_rotor_formalism} and apply it to an $L_x\times L_y$ cluster, examplified in Fig.~\ref{fig:4_x_4_lattice} by a $4\times 5$ cylindrical lattice.

\subsubsection{Pseudo-spin sector}
\label{sec:chargon_sector}

\begin{figure}[h!]
  \centering
    \includegraphics[scale=0.5]{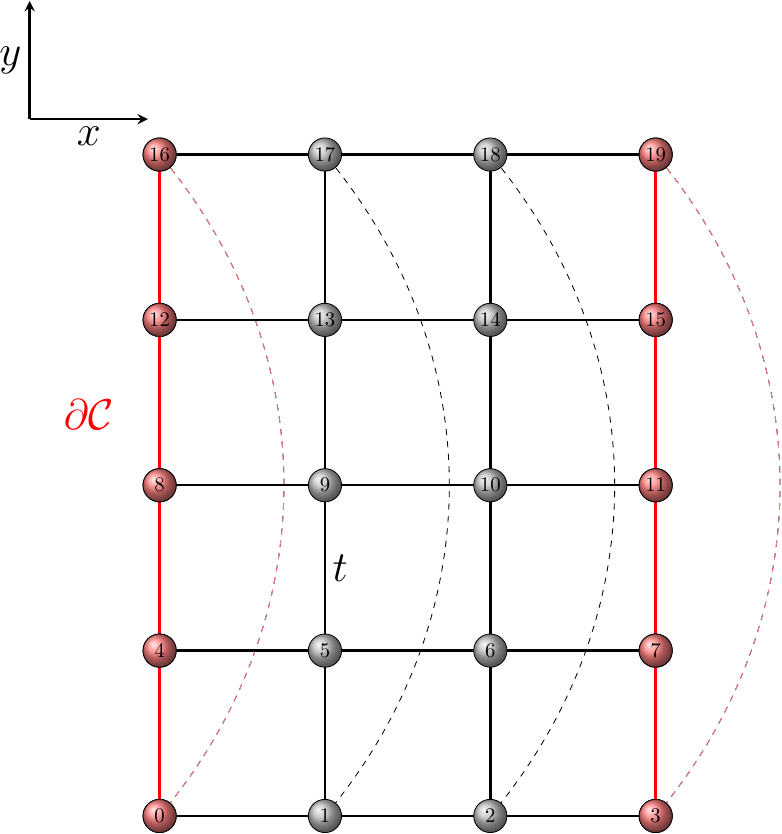}
      \caption{\textbf{Illustration of a $4\times 5$ cylindrical cluster} embedded self-consistently in a mean-field environment. The black dashed lines represent the NN periodic hopping along $L_y$. The red sites and edges indicate the open sides of the cylindrical cluster. $t$ represents the NN hopping. $\mathbf{a}=\left(0,1\right)a$ and $\mathbf{b}=\left(1,0\right)a$ are the Bravais lattice vectors tiling the full square lattice, with the lattice spacing $a$ set to unity ($a=1$).}
  \label{fig:4_x_4_lattice}
\end{figure}

The interacting pseudo-spin Hamiltonian $\hat{H}^{s}_{tUVJ}$~\eqref{eq:decoupled_equations_Ham_theta} contains terms favoring various ordering tendencies. The competition between the first two terms is characterized by the mean-field order parameter $\Phi$
\begin{align}
\label{eq:phi_order_parameter}
\Phi = \langle \hat{S}^{\pm}_i\rangle_s,
\end{align}
which is purely local.
To study a potential ordering with respect to $\Phi$, we need to allow for symmetry breaking, which in turn is made possible by a cluster mean-field approach. 
In this approach, we embed a finite-size cluster in an environment: while the cluster $\mathcal{C}$ is solved exactly, the cluster-environment coupling terms (on the boundary $\partial \mathcal{C}$, see Fig.~\ref{fig:4_x_4_lattice}) are handled at the mean-field level, leading to 

\begin{widetext}
\begin{align}
\label{eq:4_sites_square_impurity}
\hat{H}_{tUVJ}^{s} &\simeq -2t\chi\sum_{\langle ij\rangle\in\mathcal{C}}\left(\hat{S}^{-}_i\hat{S}^{+}_j + \text{H.c.}\right) + \frac{U}{2}\sum_{i\in\mathcal{C}}\hat{S}^{z}_{i}\left(\hat{S}^{z}_{i}-1\right) + V\sum_{\langle ij\rangle\in\mathcal{C}}\left(\hat{S}^{z}_{i}-1\right)\left(\hat{S}^{z}_{j}-1\right) - \mu_s\sum_{i\in\mathcal{C}}\hat{S}^z_{i} \notag\\
&\phantom{=}+ J\sum_{\langle ij\rangle\in\mathcal{C}} \ \langle\hat{\mathbf{S}}^f_i\cdot\hat{\mathbf{S}}^f_j\rangle_f \ \hat{S}_i^-\hat{S}^+_i\hat{S}_j^-\hat{S}^+_j - 2t\chi\Phi\sum_{i\in\partial\mathcal{C}}\left(\hat{S}^{-}_i + \text{H.c.}\right)
\end{align}
\end{widetext}
where we assumed that NN $\chi_{ij} \approx \chi$ for NN bonds Eq.~\eqref{eq:decoupled_equations_Ham_theta} and that $\chi$ is an average over all NN bonds. The set of cluster sites is denoted $\mathcal{C}$, while the set of sites located on the borders (open sides) is denoted $\partial\mathcal{C}$ ($\partial\mathcal{C}\subset\mathcal{C}$). In Fig.~\ref{fig:4_x_4_lattice}, we show an example of a $4\times5$ cylindrical cluster, whose borders $\delta \mathcal{C}$ are marked in red. We will still use the notation $L_x\times L_y$ to describe the lattice sizes considered. Equation~\eqref{eq:4_sites_square_impurity} can be decomposed into an intracluster part (the first five terms) and an intercluster part (the last term). It describes the pseudo-spin sector of the $t$–$U$–$V$–$J$ model as an effective quantum spin-1 XXZ model subject to a magnetic field, featuring both a transverse field (the $-2t\chi\Phi$ term) and a second-order uniaxial anisotropy (the $U$ term).

In Appendix~\ref{app:Hubbard_slave_rotor}, we illustrate some results from solving a $2\times2$ cluster embedded in a homogeneous mean-field using the slave-spin-1 theory.

Using the correspondence (Eq.~\ref{eq:original_fermion_mappings}) between pseudo-spin states and physical states, we see that a nonzero $\Phi$ means that pseudo-spins can flip in the perimeter of the cluster and hence the electronic charge can fluctuate in the bulk (correlated metal): this corresponds to a Fermi-liquid (or metallic) phase.
In contrast, the $\Phi=0$ phase corresponds to the Mott-insulating phase and a charge gap. 
When doping the cluster with holes ($\delta>0$), the bad metal (doped Mott-insulator) phase is characterized by a low, but finite (of the order of $10^{-2}$ at most), value of $\Phi$.

Let us now turn to the details of the numerical solution of this model. To initiate the calculations, $\Phi$ is set to some arbitrary, but small value. Eq.~\eqref{eq:4_sites_square_impurity} is cast into the basis $\{\ket{S_i^{z}}_s\}$ and solved directly using DMRG~\cite{RevModPhys.77.259,PhysRevLett.69.2863}, making use of the utilities provided by ITensor~\cite{10.21468/SciPostPhysCodeb.4}. The pseudo-spin density (magnetic moment) $\langle \hat{S}_i^{z}\rangle_s$ and $\Phi$ \eqref{eq:phi_order_parameter} are updated, while searching for a $\mu_{s}$ that matches the hole doping $\delta$ (see Eq.~\eqref{eq:constraint_eqs_2}) at each iteration. Upon convergence of the parameters in the pseudo-spin sector, the $B_{ij}$ correlation functions defined in Eq.~\eqref{eq:decoupled_equations_Ham_f} are computed within the pseudo-spin ground state for the NN bonds, $B_{\langle i,j\rangle} = \langle \hat{S}^{\pm}_i\hat{S}^{\mp}_j\rangle_{s}$. The bonds that feature in the $B$'s are defined on clusters similar to that depicted in Fig.~\ref{fig:4_x_4_lattice}. Also, since we are keeping all the bond-resolved and site-resolved information about the lattice throughout the procedure, $B_{\langle i,j\rangle}$ is a symmetric real matrix of size $2(L_x-\frac12)L_y\times2(L_x-\frac12)L_y$. The values of $B_{\langle ij\rangle}$ are then passed to the pseudo-fermion sector once $\mu_{s}$ and $\Phi$ have converged. In addition to the NN 2-point correlation functions, the 4-point correlation function appearing in Eq.~\eqref{eq:decoupled_equations_Ham_f} $\propto \langle\hat{S}_i^-\hat{S}^+_i\hat{S}_j^-\hat{S}^+_j\rangle_s$ is also computed in the pseudo-spin sector to be passed to the pseudo-fermion sector. The pseudo-spin sector is depicted pictorially in Fig.~\ref{fig:flow_chart_scheme} in connection with the pseudo-fermion sector, which we detail next.

\subsubsection{Pseudo-fermion sector}
\label{sec:spinon_inh_sector}

We now focus on the pseudo-fermion Hamiltonian \eqref{eq:decoupled_equations_Ham_f}.
It is an effective Heisenberg model of spin-1/2 pseudo-fermions that can hop around.
Because of the quartic $J$ term, this Hamiltonian is prohibitively difficult to solve at large cluster sizes.
Inspired by Ref.~\cite{PhysRevB.108.035139}, we decouple this term using inhomogeneous mean-field theory, namely we approximate (as detailed in Appendix~\ref{sec:inhomogeneous_MF})
\begin{align}
\label{eq:simplified_pseudo_fermion_term}
\hat{\mathbf{S}}^f_i\cdot\hat{\mathbf{S}}^f_j \simeq \sum_{\sigma, \sigma'}\alpha_{\sigma,\sigma^\prime}^{ij}\hat{f}^\dagger_{i,\sigma}\hat{f}_{j,\sigma^\prime},
\end{align}
with $\alpha$ a tensor containing the mean-field parameters.

This allows us to compute the 2-point correlation function $\chi_{ij}$ and 4-point correlation function $\langle\hat{\mathbf{S}}^f_i\cdot\hat{\mathbf{S}}^f_j\rangle_f$. 
Note that the correlators calculated here are confined to clusters like the one defined in Fig.~\ref{fig:4_x_4_lattice}; it is also the case for the correlators of the pseudo-spin sector ($i,j\in\mathcal{C}$).

The 2-point correlation matrix $\chi_{ij}$ and the 4-point correlation function $\langle\hat{\mathbf{S}}^f_i\cdot\hat{\mathbf{S}}^f_j\rangle_f$ are then injected back into the pseudo-spin sector, to carry on with the self-consistency computation. Note that one could make use of correlations beyond NN if the cluster permits and long-range interactions ought to be considered. Importantly, throughout the self-consistent cluster mean-field procedure, the spatial resolution of the correlators $\chi_{ij}$ and $B_{ij}$, as well as higher-order correlators, is retained/preserved. In Appendix~\ref{app:Hubbard_slave_rotor}, we show results of the self-consistent slave-spin-1 cluster theory assuming homogeneous slave-spin-1 mean fields.

In all the results below, the NN hopping $t$ serves as an energy reference for the other parameters. The calculations are carried out at zero temperature.

\section{Self-consistent mean-field solution}
\label{sec:self_consistent_mf_solution}

We now present the results obtained by implementing the self-consistent slave-spin cycle described by Fig.~\ref{fig:flow_chart_scheme}. We initialize this cycle by randomly setting the values of the spatially resolved correlation functions $\chi_{ij}$'s and the 4-point correlation function $\langle \mathbf{S}^f_i\cdot\mathbf{S}^f_j\rangle_f $ (Eq.~\eqref{eq:expectation_val_SiSj}) (while preserving their symmetric structure).
We do not use any mean-field pinning field. The solutions upon convergence do not significantly depend on the initial point, even when starting from an initial random state.

\begin{figure*}[htb!]
  \centering
    \includegraphics[width=\textwidth]{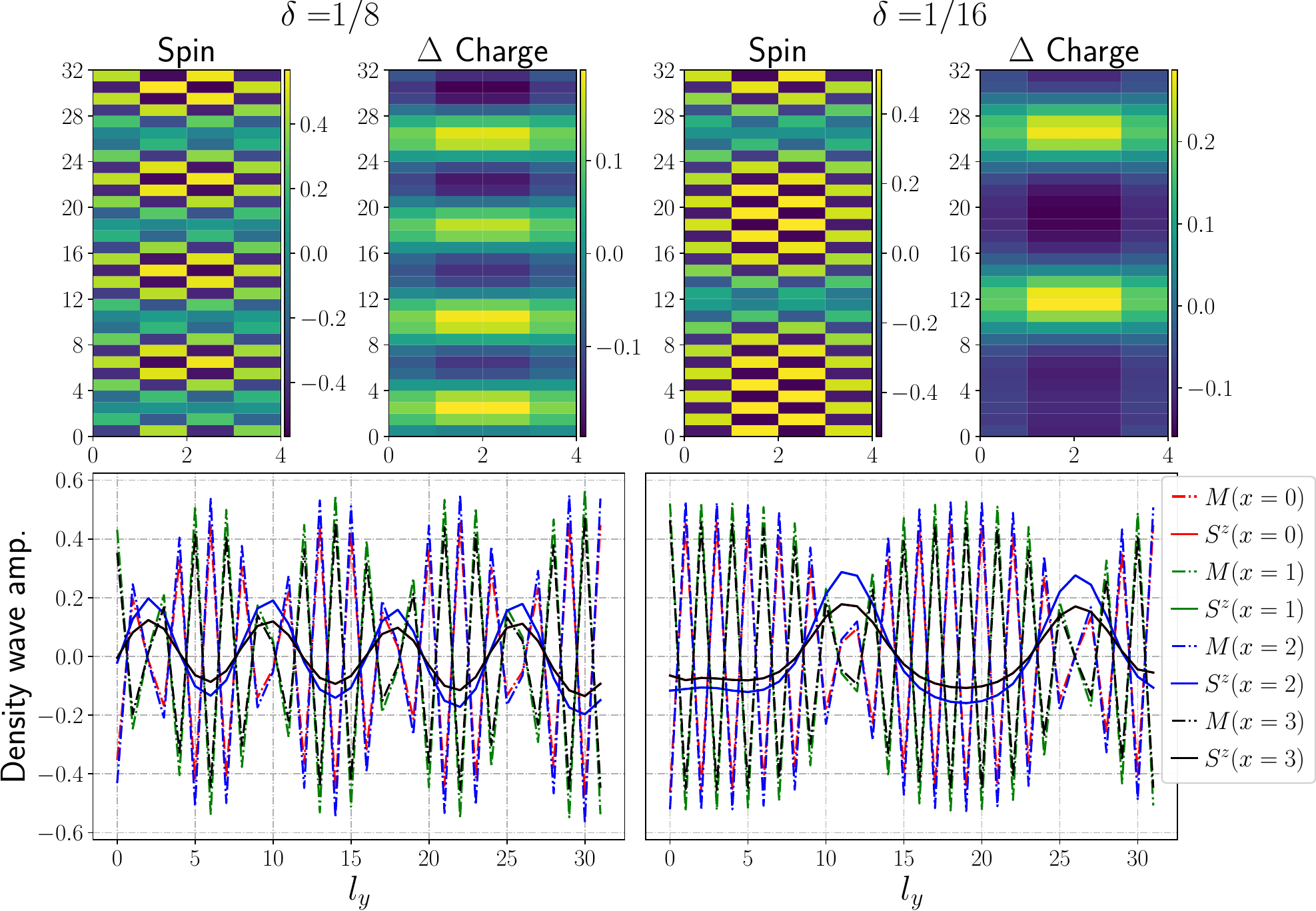}
      \caption{\textbf{Real-space signatures of the SDWs and CDWs as a function of doping at $J=0.2$ and $V=0$.} Upper left panel: magnetization showing staggered AFM ordering in the pseudo-fermion sector (left inset) and pseudo-spin-1 modulations about the average magnetic moment showing CDWs (right inset)---the yellow (blue) color relates to surplus of holes (electrons) with respect to half-filling. The value of $\delta=1/16$ and $U=1.5$. Upper right panel: same as upper left panel, although for values of $\delta=1/8$ and $U=2$. Bottom left panel: pseudo-fermion magnetization $M$ along strips $l_x$ of 32 sites (dotted curves) and spin-1 projections $S^z$ (solid curves) for the same set of parameters as the panel above ($\delta=1/16$ and $U=1.5$). Bottom right panel: same data presented as that of the bottom left panel, although for parameters $\delta=1/8$ and $U=2$. For readability, the charge oscillations are amplified by a factor $10$ for both panels. We remind here that DMRG is only employed for the spin-1 XXZ transverse-field model that represents the charge sector.The calculations were carried out on a $4\times 32$ lattice with periodic boundary conditions along the longest side $L_y$.}
  \label{fig:CDW_SDW}
\end{figure*}

\subsection{Stripes}
\label{sec:stripes}

\begin{figure*}[htb!]
  \centering
    \includegraphics[width=\textwidth]{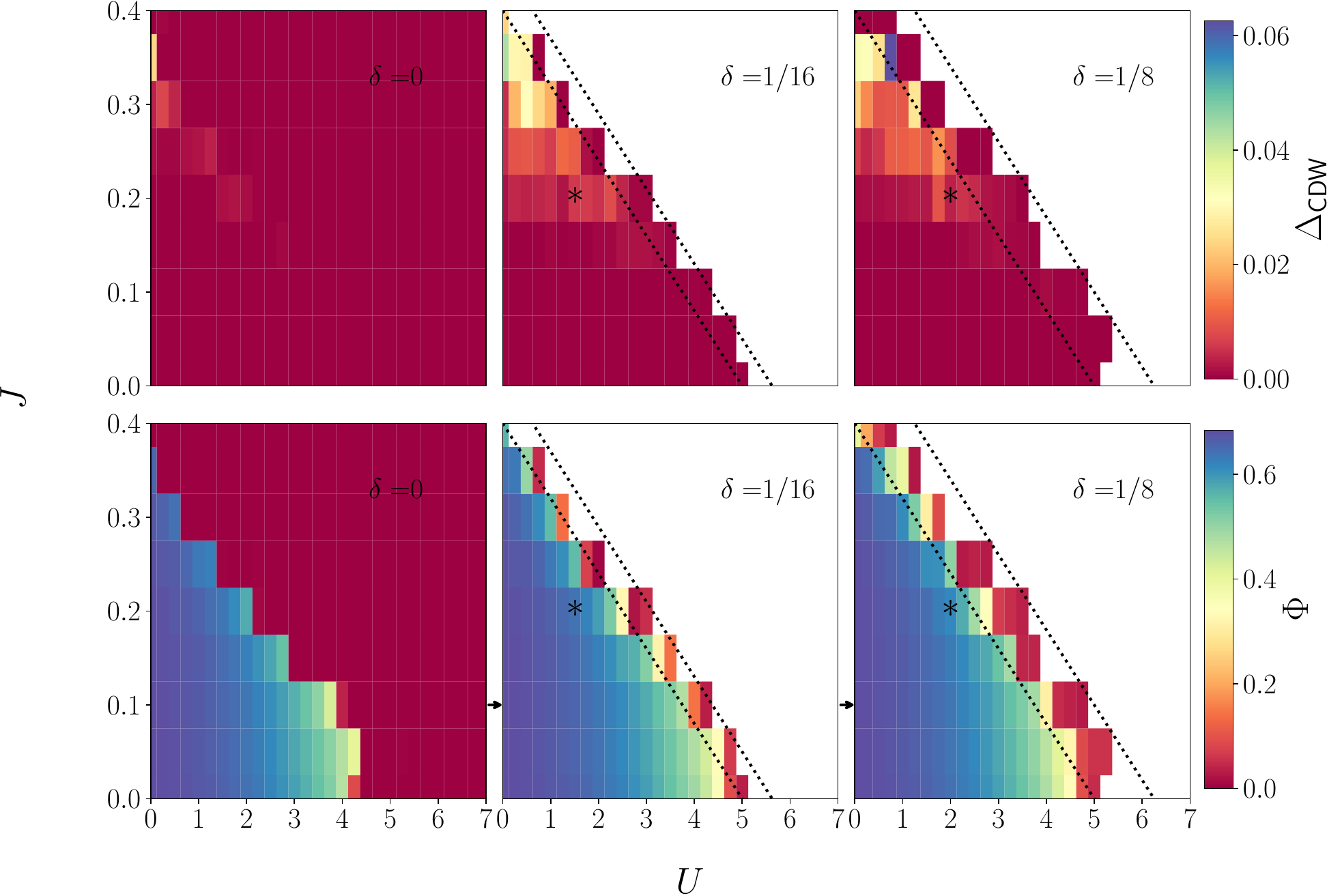}
      \caption{\textbf{CDW ($\Delta_{\text{CDW}}$) and metallic ($\Phi$) order parameters vs $J$ and $U$ for various hole dopings.} The value of the NN interaction $V=0$. Top panels: $\Delta_\text{CDW}$ for $\delta=0,1/16,1/8$, going from left to right. Bottom panels: $\Phi$ for $\delta=0,1/16,1/8$, going from left to right. The top (bottom) color bar shows the intensity of $\Delta_\text{CDW}$ ($\Phi$). The white regions display the parameter spaces where the slave-spin method fails to converge. The region between the diagonal dotted lines roughly encloses the region where phase separation occurs. The values of $\Phi$, when converging never reach numbers below $\sim 10^{-4}$. The calculations were carried out on a $4\times 32$ lattice with periodic boundary conditions along the longest side $L_y$. The black asterisks mark the data points shown in Fig.~\ref{fig:CDW_SDW}, while the black arrows mark that shown in Fig.~\ref{fig:Phi_vs_J_U_zoom_J_0p1}.}
  \label{fig:Phi_vs_J_U}
\end{figure*}


In Fig.~\ref{fig:CDW_SDW}, we show the real-space spin and charge density distributions obtained at convergence of the self-consistent loop for two different doping levels $\delta$ and interaction values $U$, for clusters of size $4\times32$.
The spin density (obtained from the solution of the pseudo-fermion sector) displays an alternation of large horizontal AFM domains with narrow horizontal domains with no AFM character. Meanwhile, the charge density (obtained from the solution of the pseudo-spin sector) displays horizontal domains with excess charge (in blue) and horizontal domains with charge depletion (in yellow).

We can clearly see that the holes aggregate (yellow colors in the top right insets) where the AFM magnetization amplitude decreases.
Conversely, an enhanced AFM magnetization also coincides with a larger density of electrons  (blue colors in the top right insets).
In the bottom panels of Fig.~\ref{fig:CDW_SDW}, we display both the spatially-resolved magnetization and charge density (pseudo-spin magnetization) for cuts taken along $L_x$, for the corresponding panels shown just above.
The phased-in behavior of the charge and spin stripes is even more striking under this point of view: where holes aggregate (crests in $S^z$), the spin-1/2 magnetization is lowest (troughs in $M$) for all $x\in\left[0,L_x-1\right]$.

The periodicity of the stripes depends on the hole doping level.
In Fig.~\ref{fig:CDW_SDW}, we observe that doubling the doping roughly halves the wavelength $\lambda^y_{\text{CDW}}$: going from $\delta = 1/16\to 1/8$, the wavelength of oscillations of both CDW and SDW stripes halves, going from  $\sim14-15$ (resp. $\sim28-30$) sites to  $\sim7-8$ (resp. $\sim15-16$) sites for the  CDWs (resp. SDWs). This periodicity corresponds to that of filled stripes: $\lambda^y_{\text{CDW}(\text{SDW})}=\frac{1(2)}{\delta}$.

Let us now more systematically explore the characteristic of the stripes across the $J-U$ plane.
Given that all the clusters we studied have a commensurate electron filling (see Appendix~\ref{sec:cluster_geometry_size}), both SDWs and CDWs share the same phase, and therefore, to map the stripes in the phase diagram, we fit the CDWs with the function 
\begin{align}
\label{eq:stripes_fit_function}
f(\mathbf{x};\mathbf{Q},\Delta,\phi,b) = \Delta_{\text{CDW}}\cos{\left(\mathbf{Q}_{\text{CDW}}\cdot\mathbf{x}_i + \phi\right)} + b,
\end{align}
where $\Delta_{\text{CDW}}$ denotes the amplitude of the CDWs, $b$ some small offset close to $0$, and $\mathbf{x}_i$ the position of the lattice site. 

In the top panels of Fig.~\ref{fig:Phi_vs_J_U}, we display in the $U-J$ plane the CDW amplitude $\Delta_{\text{CDW}}$ for $4\times32$ clusters obtained by fitting the charge stripes with Eq.~\eqref{eq:stripes_fit_function}, for $\delta=\{0,1/16,1/8$\}.
At half-filling (top left panel), stripe physics is absent; the few points at large $J$ values that correspond to noticeable finite $\Delta_{\text{CDW}}$ are irregular inhomogeneous charges density modulations. At large $J$, the variational landscape seems rugged and full of local minima. The stripes are only significant at finite doping within the slave-spin-1 method.

When doping to $\delta=1/16$, the region in $J-U$ space wherein $\Delta_{\text{CDW}}>0$ lies above $J\geq0.15$ (same for $\delta=1/8$). Clearly, a finite, but not too large, value of $J$ is required to have well-defined stripes. For these values of $J$, at larger $U$, $\Delta_{\text{CDW}}$ starts fading away, especially in the poorly conducting region of the phase diagram (red-yellow regions in bottom panels of Fig.~\ref{fig:Phi_vs_J_U}), where the stripes are vanishingly small. Doping away from half-filling increases the range of $U$ over which stripe signatures are observed. When $U$ is large, in the poor conducting regions delimited by the dotted black diagonal lines, the pseudo-spin moments concentrate within thin strips; in the bad insulating phase at finite doping, the stripes are much less spread out and AFM domains form. Once the system is in the doped-insulating phase (red-yellow regions in bottom panels of Fig.~\ref{fig:Phi_vs_J_U}), the pseudo-spin moments in the pseudo-spin sector saturate to $S^{z}_i = 0$ for most sites $i$, thereby concentrating in space the regions where $\langle\hat{S}_i^{z}\rangle_{s}\neq0$---pseudo-spins order ferromagnetically and form domains (except at half-filling). Beyond this poorly conducting regions, going into the white region, the slave-spin-1 method can no longer enforce the constraints~\eqref{eq:constraint_eqs_2} and does not converge.
In Appendix~\ref{sec:cluster_geometry_size}, we show results similar to Fig.~\ref{fig:CDW_SDW} for a cluster of size $4\times16$.

\begin{figure}[htb]
  \centering
    \includegraphics[scale=0.4]{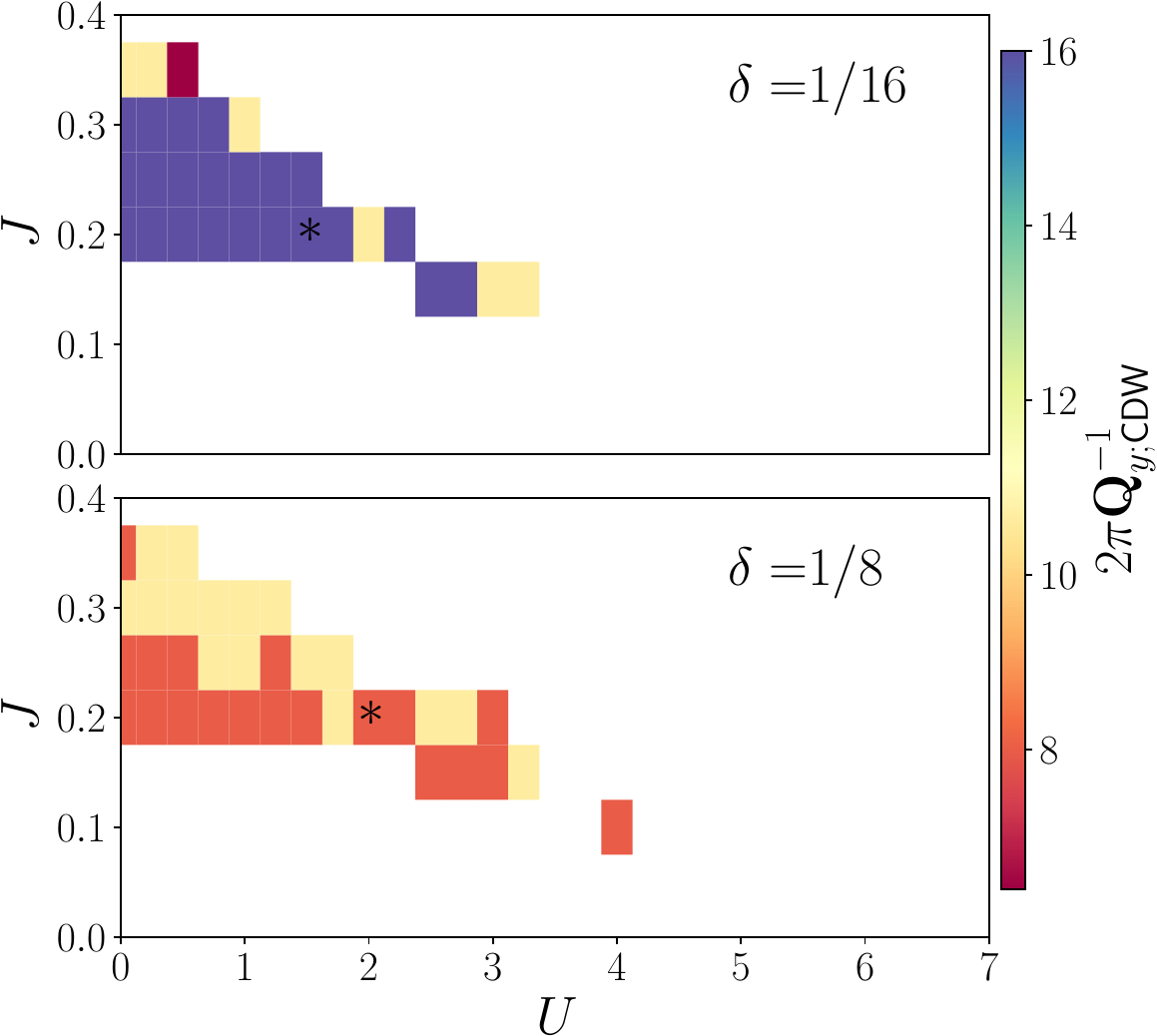}
      \caption{\textbf{wavelength of the CDW order parameter $\mathbf{Q}_{\text{CDW}}$} vs $U$ and $J$ for hole dopings $\delta=\{1/16,1/8\}$, and nearest-neighbor interaction $V=0$. The clusters sizes are $4\times32$. The color map shows the wavelength of the stripes in terms of amount of lattice sites, namely $2\pi\mathbf{Q}_{y;\text{CDW}}^{-1}$. A threshold of $\Delta_{\text{CDW}}\geq 1\times10^{-3}$ was put, and that explains why the white region widened when compared to Fig.~\ref{fig:Phi_vs_J_U}. The black asterisks mark the data points shown in Fig.~\ref{fig:CDW_SDW}.}
  \label{fig:Q_CDW_vs_J_U}
\end{figure}

Let us finally study the $(U,J)$ dependence of the CDW ordering wavevector. 
In Fig.~\ref{fig:Q_CDW_vs_J_U}, we plot $\lambda^y_{\text{CDW}}\equiv2\pi\mathbf{Q}_{y;\text{CDW}}^{-1}$, that is the CDW wavelength resolved in amount of sites along the $y$-axis (see Fig.~\ref{fig:4_x_4_lattice}), in the $J-U$ plane for both $\delta=1/16$ (top panel) and $\delta=1/8$ (bottom panel). $\lambda^y_{\text{CDW}}$ amounts to calculating the number of sites separating two consecutive crests or troughs in the CDWs.
One can clearly see from comparing the two dopings that $\lambda^y_{\text{CDW}}$ decreases upon doping away from half-filling. It seems that $\lambda^y_{\text{CDW}}$ does not vary much in the $J-U$ plane at $\delta=1/8$, as opposed to $\delta=1/16$. Indeed, for $\delta=1/8$, $\lambda^y_{\text{CDW}}\simeq 7-11$ sites, while for $\delta=1/16$, $\lambda^y_{\text{CDW}}\simeq 11-16$ sites.
We nevertheless observe that the period of the stripes roughly doubles when halving the hole content for $J\leq0.25$, consistent with Refs.~\cite{PhysRevB.108.205154,10.21468/SciPostPhys.7.2.021,doi:10.1126/science.adh7691}. The black asterisks in Fig.~\ref{fig:Q_CDW_vs_J_U} indicate where the data shown in Fig.~\ref{fig:CDW_SDW} sit.

It is important to stress that the stripes do not stabilize if the cluster is completely open, and that at half-filling, no significant CDWs nor SDWs are observed (see top left panel in Fig.~\ref{fig:Phi_vs_J_U}); in this case, we only have Néel AFM in the pseudo-fermion sector and somewhat homogeneously distributed moments in the pseudo-spin sector giving $\langle\hat{S}_i^{z}\rangle_{s}=0$. For the few points at half-filling and large $J$ where $\mathbf{Q}_{\mathrm{CDW}} \neq \mathbf{0}$, the charge and spin modulations are highly inhomogeneous and should not be considered as stripes.

In the correlated metal (see below for more details), we can assess the robustness of the stripes obtained there, having run many simulations starting out with some random initialization of the inhomogeneous mean fields in the pseudo-fermion sector (see Appendix~\ref{sec:cluster_geometry_size} for other cluster sizes). At fixed $U$, stripe solutions initialized with independent runs can differ by a phase factor $\phi$, while the wavelength of oscillations $\lambda^y_{\text{CDW}}$ remains unchanged. We note that since there is a mean-field decoupling between the charge and spin sectors, the effective values of $U$ correspond to higher ones in estimations from more sophisticated strongly-correlated methods like DMFT, since the electronic charge and spin screenings are underestimated here~\cite{PhysRevB.87.125149,Vilk_1997,PhysRevB.107.245137,PhysRevB.103.104415,PhysRevB.108.035139,PhysRevB.104.245127}.

\subsection{Metal-insulator transition}
\label{sec:metal_insulator_phase_transition}

The transverse field $\Phi$ (Eq.~\eqref{eq:phi_order_parameter}) of the XXZ model serves as an order parameter that determines whether the system is in a metallic or an insulating phase. Its role is akin to the quasiparticle weight, as discussed in Ref.~\cite{PhysRevB.70.035114,PhysRevB.66.165111}. A finite $\Phi$ indicates a metallic state, whereas $\Phi=0$ (within numerical precision) signals a Mott insulator. Upon doping, the Mott-insulating phase is absent, resulting instead in a small but finite $\Phi$ characteristic of a bad insulating regime (doped Mott insulator).

\begin{figure}[htb!]
  \centering
    \includegraphics[scale=0.3]{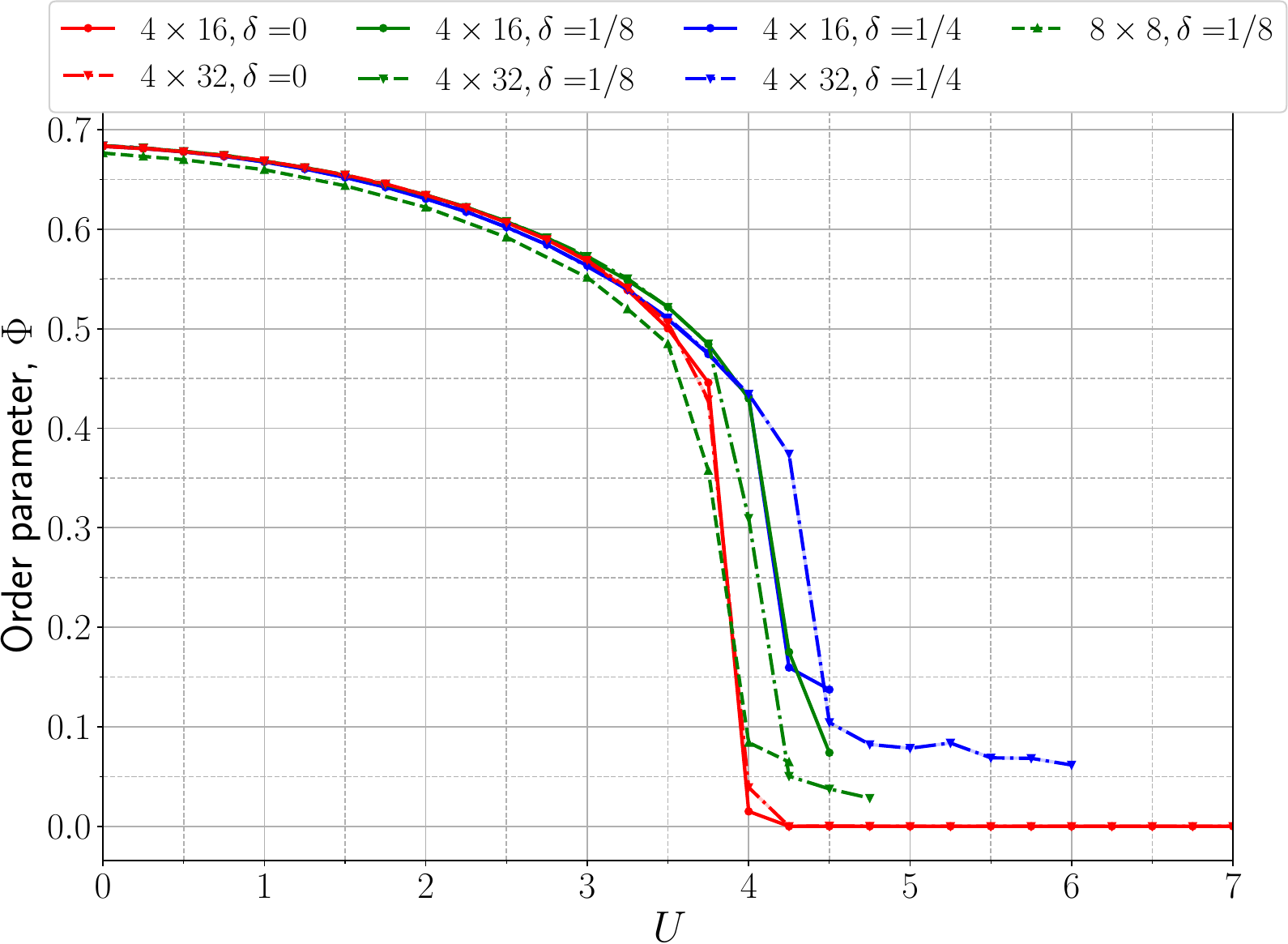}
      \caption{\textbf{Metallic order parameter $\Phi$ vs $U$ for hole dopings} $\delta=\{0,1/8,1/4\}$, $J=0.1$ and nearest-neighbor interaction $V=0$. The clusters sizes are $8\times8$, $4\times 16$ and $4\times32$, and are periodic boundary conditions along the longest side $L_y$. The line styles and markers are shared among the same cluster sizes, while the colors are shared among the same dopings.}
  \label{fig:Phi_vs_J_U_zoom_J_0p1}
\end{figure}

Fig.~\ref{fig:Phi_vs_J_U_zoom_J_0p1} displays the evolution of $\Phi$ as a function of the local interaction $U$ for several hole dopings $\delta=\{0,1/8,1/4\}$, at a fixed spin exchange coupling $J=0.1$, $V=0$  and cylinders of sizes $4\times16$, $4\times32$ and $8\times 8$. 

We see that a sharp transition of $\Phi$ to $0$ occurs only at half-filling, while doping away from half-filling smoothens this sharp metal-insulator drop since $\Phi$ remains finite. Clearly, the larger the hole doping, the better the system conducts across a wider range of $U$. It is also apparent that changing the system size does not significantly affect $\Phi$ at any finite doping:
regardless of the cluster size, all the curves in Fig.~\ref{fig:Phi_vs_J_U_zoom_J_0p1} seem to converge to the same value of $\Phi$ as $U\to0$ ($\Phi\simeq 0.7$), and that value is also found in Ref.~\cite{PhysRevB.76.195101} for the truncated slave-rotors.

We now investigate the variations of $\Phi$ across the $(U, J)$ plane for the cluster size $4\times32$ in the bottom panels of Fig.~\ref{fig:Phi_vs_J_U} for different fillings (half-filling ($\delta=0$), $\delta=1/16$ and $\delta=1/8$).
Note that the data presented in Fig.~\ref{fig:Phi_vs_J_U_zoom_J_0p1} for the $4\times32$ cluster corresponds to the $J=0.1$ lines from the bottom panels of Fig.~\ref{fig:Phi_vs_J_U} at corresponding $\delta$ and flattening it out along $U$ (see black arrows).
For all dopings, increasing the value of $J$ reduces the critical $U_c$ above which $\Phi\to0$, corresponding to a bad metal at finite doping or an insulating state ($\Phi=0$) at half-filling.
In other words, stronger spin fluctuations shift the charge gap opening to lower values of $U$. Hole-doping suppresses the Mott transition, with $\Phi\gtrsim1\times10^{-4}$ in the bad insulating phase (red region). For most values of the spin exchange coupling, the metallic region seems to stretch out to larger values of $U$ (light green and violet region), before taking values associated to a poor conducting state. Indeed, increasing the hole doping pushes upwards in $U$ the bad insulator phase, especially since there are fewer electrons per site on average and therefore one needs a stronger value of $U$ to disadvantage doublons characterized by $\langle\hat{S}^z\rangle_{s}=-1$ (the kinetic energy is large compared to the potential energy).

In the doped regime, the convergence of the self-consistency becomes difficult in the large $(U,J)$ regime. There, the value of $\Phi$ starts oscillating around small values in the poorly insulating state and one needs to start the simulations from previously converged weaker $U$ values, while ensuring that the correlation functions from one iteration to the next get mixed to a great extent ($\gtrsim 80\%$ mixing). At even larger values of $U$ and $J$, away from half-filling, the slave-spin self-consistency breaks down and does not converge with the sought $\delta$. This is shown by the white regions of Fig.~\ref{fig:Phi_vs_J_U}. In Appendix~\ref{sec:cluster_geometry_size}, we show results similar to Fig.~\ref{fig:Phi_vs_J_U} for a cluster of size $4\times16$.

\subsection{Effect of $V$}
\label{sec:V_effect}

We now study the effect of having a nonzero NN Coulomb interaction $V$ on the doped Mott-insulating phase as well as on the spin and charge stripes.

\begin{figure}[h!]
  \centering
    \includegraphics[width=\columnwidth]{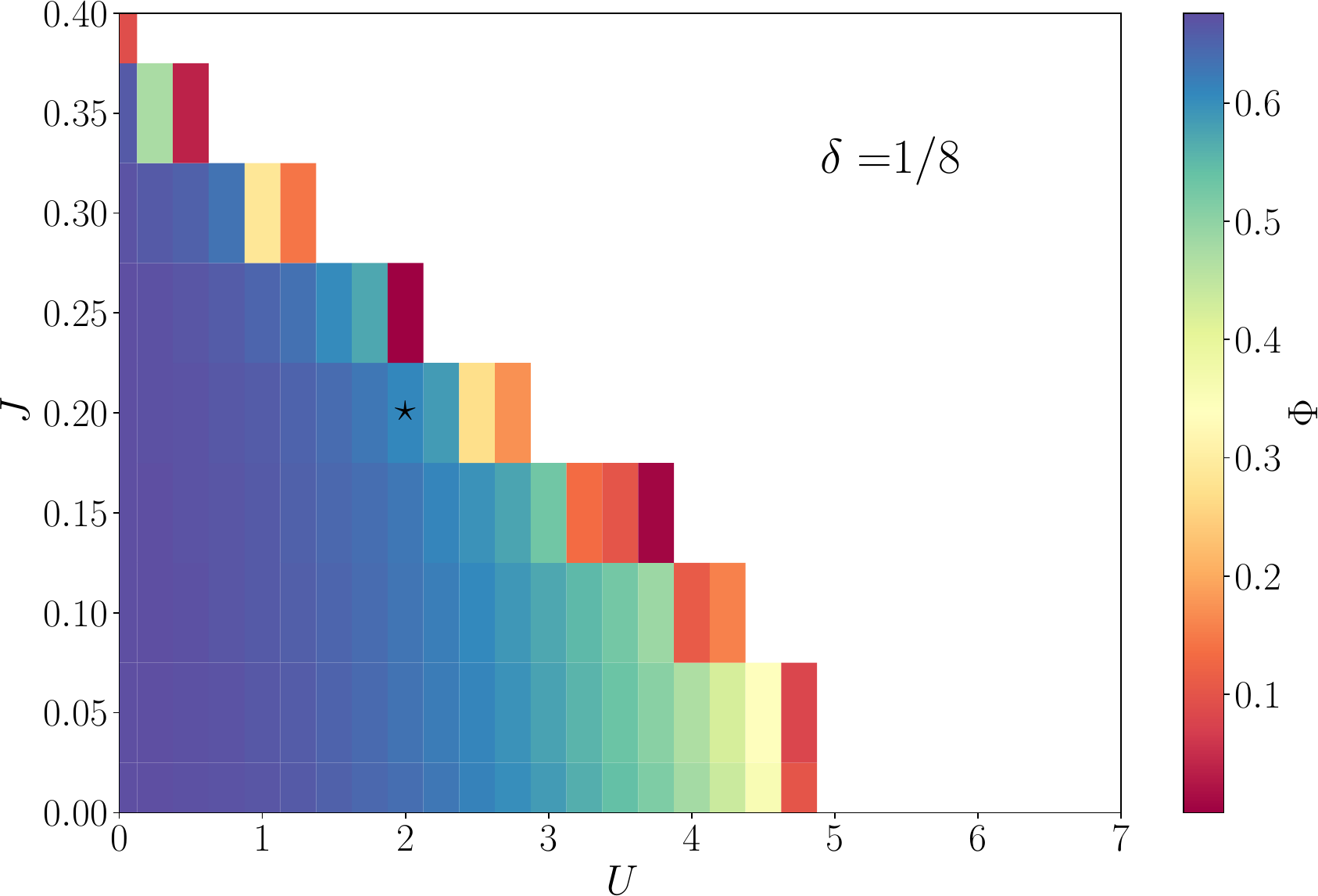}
      \caption{\textbf{Metallic order parameter $\Phi$ vs $J$ and $U$ for hole doping $\delta=1/8$} and nearest-neighbor interaction $V=0.04$. The color bar shows the intensity of $\Phi$. The calculations were carried out on a $4\times 32$ lattice with periodic boundary conditions along the longest side, which is $L_y$. The white region illustrates the phase space where convergence of the slave-spin-1 method fails. The black star relates in the $J-U$ plane Fig.~\ref{fig:CDWs_SDWs_V_finite}.}
  \label{fig:Phi_vs_J_U_V_finite}
\end{figure}

In Fig.~\ref{fig:Phi_vs_J_U_V_finite}, we show a color map of $\Phi$ in the parameter space of $J$ and $U$, for $V=0.04$. This color map can be directly compared to Fig.~\ref{fig:Phi_vs_J_U} (top right panel) for the case where $V=0$. As a first observation, the non-white region associated to a correlated metal is shrunk overall. The system turns into a doped Mott insulator at lower values of $J$ (fixed $U$) and $U$ (fixed $J$). Adding some extra NN interaction $V$ indirectly increases $U$ by penalizing electrons---pseudo-spin-1 with projections $\langle\hat{S}^z\rangle_s = 0,-1$---to sit next to each other in neighboring sites. Moreover, from looking at Eq.~\eqref{eq:decoupled_equations_Ham_theta}, one can see that the last term related to $V$ induces extra AFM correlations in the pseudo-spin sector, which competes with the FM in-plane term (first term).

\begin{figure}[h!]
  \centering
    \includegraphics[scale=0.45]{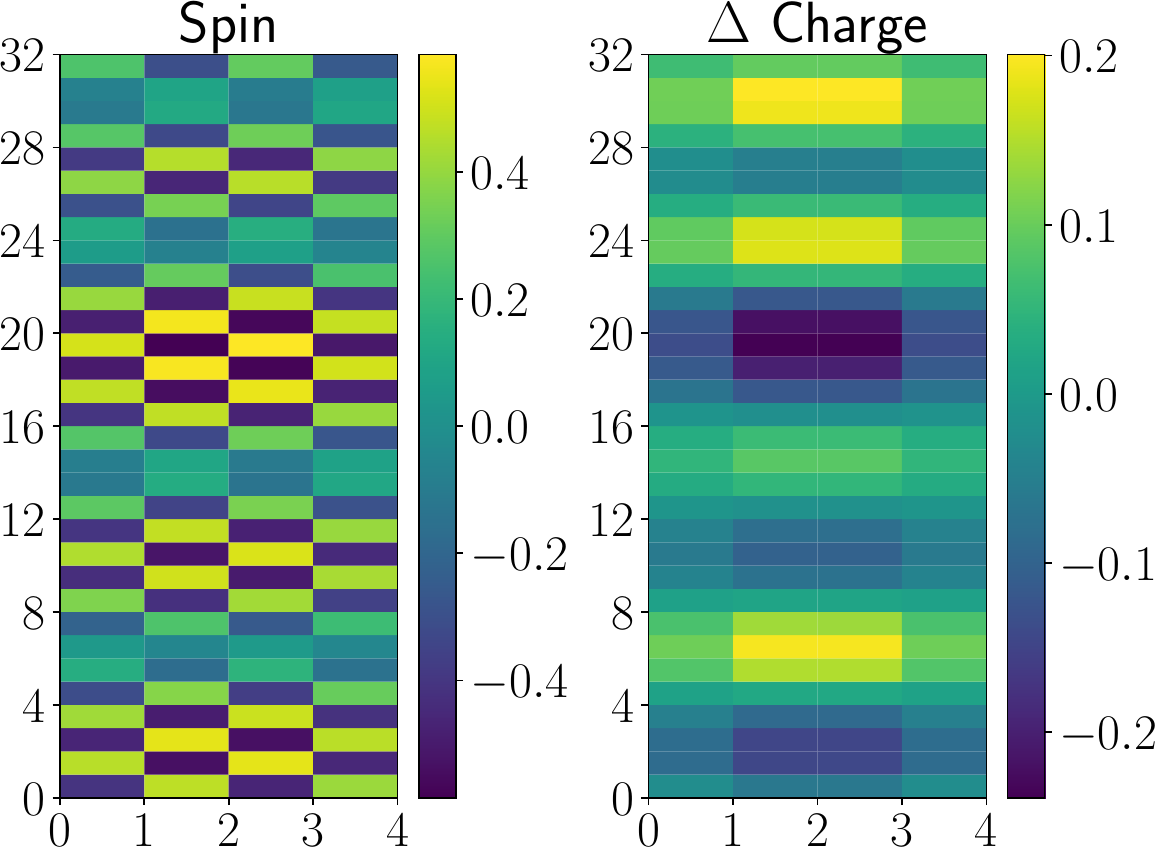}
      \caption{\textbf{SDWs and CDWs at a finite $V$}. The NN interaction $V=0.04$, $J=0.2$ and $U=2$. The hole doping is $\delta=1/8$. The only difference with the top left panel of Fig.~\ref{fig:CDW_SDW} is $V$ here. Left panel: Magnetization as a function of the lattice sites. Right panel: pseudo-spin-1 magnetic moment variations about the average for all cluster sites. The charge oscillations are amplified by a factor $10$, like in Fig.~\ref{fig:CDW_SDW}. In Fig.~\ref{fig:Phi_vs_J_U_V_finite}, a black star indicates where this parameter set lies in the $J-U$ plane.}
  \label{fig:CDWs_SDWs_V_finite}
\end{figure}

We now turn to Fig.~\ref{fig:CDWs_SDWs_V_finite}, where the effect of a finite $V=0.04$ onto the CDWs and SDWs is looked at. To do so, we employ the same set of parameters as those used for the $V=0$ case shown in Fig.~\ref{fig:CDW_SDW} (top left panel), namely $U=2$, $J=0.2$ and $\delta=1/8$. As seen in Fig.~\ref{fig:CDW_SDW}, top left panel, the stripes amount to four periods. However, the stripes here with some finite $V$ are a bit less regular. The SDWs show similar pattern and intensity when looking at the color map for both $V=0.04$ and $V=0$. The amplitudes of the CDWs are increased when compared to the case $V=0$. This is due to the fact that, on top of the AFM modulations that imprint some modulation in the microscopic couplings of the pseudo-spin model, NN Coulomb interactions tend to squeeze together electrons and holes on equal footing.

As seen from Fig.~\ref{fig:CDWs_SDWs_V_finite} for $V=0.04$ and Fig.~\ref{fig:Phi_vs_J_U} for $V=0$ (bottom right panel), both the CDWs and SDWs have a different phase, \textit{i.e.}, the sites along $L_y$ at which pseudo-spins have higher magnetic moments and pseudo-fermions aggregate differ, even though the stripes are roughly similar otherwise. This may however not be a consequence solely attributed to $V$, but this could be due to the facts that the ground state is degenerate with many local minima and that we started with different initial conditions: the stripy phases can arise with a different phase disregarding the values of $V$.

Note that the effects of increasing $V$ on a cluster with open boundary conditions pushes the pseudo-spin-1 to have a magnetic moment of $S^z=-1$ on the corners of the cluster (not shown), because that electrons want to occupy regions where there are less nearest neighbors.

\subsection{Dynamics between the pseudo-spin and pseudo-fermion sectors}
\label{sec:dynamics_spinon_chargon_discussion}

In this section, we explain how the interplay of the (pseudo) charge and spin degrees of freedom lead to the emergence of stripes in certain regimes (see Fig.~\ref{fig:flow_chart_scheme}).

To do that, we first study the correlations functions that are passed from one sector to the other. Recall that we deal with an AFM spin coupling, $J>0$. Owing to the AFM spin coupling, the NN spin-spin correlation functions $\langle \hat{\mathbf{S}}_i^f\cdot\hat{\mathbf{S}}_j^f\rangle_f<0$ give negative values bounded to $-1$ from below. Increasing $J$ makes $\langle \hat{\mathbf{S}}_i^f\cdot\hat{\mathbf{S}}_j^f\rangle_f\to-1$. Moreover, the $B_{ij}$ terms that renormalize the pseudo-fermion hopping energy (see Eq.~\eqref{eq:decoupled_equations_Ham_f}) remain positive and of the order $\sim1$, but get strongly suppressed in the (doped) Mott-insulating regime towards $0$. The $B_{ij}$'s diminish at fixed $U$ when $J$ grows. In a similar fashion, $\chi_{ij}$ defined in Eq.~\eqref{eq:decoupled_equations_Ham_theta} is a quantity that stays positive but depletes when $U$ increases. However, in the weak U regime, $\chi_{ij}<B_{ij}$, $\chi_{ij}$ being roughly $25\%$ the value of $B_{ij}$. $\chi_{ij}$ also decreases as a function of $J$ at fixed $U$. The term $\propto\langle\hat{S}_i^-\hat{S}^+_i\hat{S}_j^-\hat{S}^+_j\rangle_s$ that appears in the pseudo-fermion sector \eqref{eq:decoupled_equations_Ham_f} on the other hand is positive and increases as a function of $U$ and $J$. When doping in holes, both the chemical potential $\mu_f$ in Eq.~\eqref{eq:decoupled_equations_Ham_f} and $\mu_s$ in Eq.~\eqref{eq:decoupled_equations_Ham_theta} are negative. A nonzero chemical potential $\mu_s$ acts like a magnetic field along the quantization axis. Doping away from half-filling maintains both $\chi_{ij}$ and $B_{ij}$ at larger values, screening the effect of the uniaxial anisotropy coming from $U$ in Eq.~\eqref{eq:decoupled_equations_Ham_theta}.

Having laid out the key behavior of the correlation functions as a function of $U$ and $J$, we focus on the pseudo-spin Hamiltonian \eqref{eq:decoupled_equations_Ham_theta} that we solve in the simulations. We recall first that $\hat{S}^{\pm}=\hat{S}^{x}\pm i\hat{S}^{y}$, such that $\hat{S}^{-}\hat{S}^{+}=\hat{S}^{2}-(\hat{S}^{z})^{2}-\hat{S}^{z}$. Therefore, $\hat{S}^{-}\hat{S}^{+}$ projects out $S^z=1$ states (empty states).
In fact, the following relation holds:

\begin{equation}
    \hat{S}_j^{-}\hat{S}_j^{+} = 2 \hat{P}_{m_{j}\neq 1},
\end{equation}
where $\hat{P}_{m_{j}\neq 1}$ is the projector on the non-empty states of site $j$. Thus, the $J$ term favors configurations with $\langle \hat{P}_{m_{i}\neq 1} \hat{P}_{m_{j}\neq 1} \rangle_s \geq 0$, namely bonds without holes; the stronger the AFM correlations, the more energetically costly it becomes to introduce holes on such bonds. In the numerics, the inclusion of $\hat{P}_{m_{i}\neq 1} \hat{P}_{m_{j}\neq 1}$, modulated by the AFM background, is a key factor for the emergence of the SDWs and CDWs.

Turning to $\hat{H}^f_{tUVJ}$~\eqref{eq:decoupled_equations_Ham_f}, the energy is minimized at fixed $B_{ij} \geq 0$ and $\langle \hat{P}_{m_{i}\neq 1} \hat{P}_{m_{j}\neq 1} \rangle_{s} \geq 0$. The first term promotes metallic conduction, since a larger $B_{ij}$ directly enhances the effective hopping amplitude. The $J$ term favors antiferromagnetic ordering whenever $\langle \hat{P}_{m_{i}\neq 1} \hat{P}_{m_{j}\neq 1} \rangle_{s} > 0$.
On the other hand, bonds containing holes, for which $\langle \hat{P}_{m_{i}\neq 1} \hat{P}_{m_{j}\neq 1} \rangle_{s} = 0$, do not contribute to magnetic correlations. The effect of doping is therefore transparent: hole doping introduces bonds with vanishing $\langle \hat{P}_{m_{i}\neq 1} \hat{P}_{m_{j}\neq 1} \rangle_{s}$, thereby locally suppressing antiferromagnetic order.

Quite remarkably, this balance between the various correlators gives rise to stripes whose qualitative behavior seem to match that of more complicated numerical methods, except that we see the stripes appearing at weaker $U$. Performing the calculations on larger lattice sizes increase significantly the simulation time. Because the stripes appear more robust and enhanced in larger lattice sizes, this gives an incentive to simulate this on a neutral atom quantum platforms, where system sizes surpass several hundreds of sites within highly-controllable systems.


\section{Comparison with direct DMRG solutions}
\label{sec:exact_solution}

The emergence of striped phases in DMRG solutions of the Hubbard model is well documented ~\cite{doi:10.1126/science.adh7691,PhysRevLett.91.136403,PhysRevLett.80.1272,doi:10.1126/science.aam7127}.
Because of the numerical cost of solving a fermionic problem like the Hubbard model with DMRG, these solutions are limited to small sizes.
In this section, we want to qualitatively compare these (virtually exact) solutions to the solutions we obtained through our slave-spin approach, which reduces a complex many-fermion problem to a self-consistently determined many-spin problem, possibly allowing to reach larger sizes (at the cost of an approximate treatment).


As a reminder, we added the $J$ term with the intention to compare direct DMRG simulations with our slave-spin-1 method, which needs it in order to exhibit stripe patterns---the slave-spin-1 treatment of the Hubbard model is unable to resolve the spin exchange physics. 

For all the direct simulations of the Hamiltonian~\eqref{eq:tUVJ_Ham}, we set the bond dimension $\chi$ to 6400, thereby ensuring that the truncation errors are roughly $1\times 10^{-5}$. For all the slave-spin-1 simulations, the bond dimension $\chi$ is caped at 300, with truncation errors hovering around $1\times 10^{-7}$.

\begin{figure}[h!]
  \centering
    \includegraphics[width=\columnwidth]{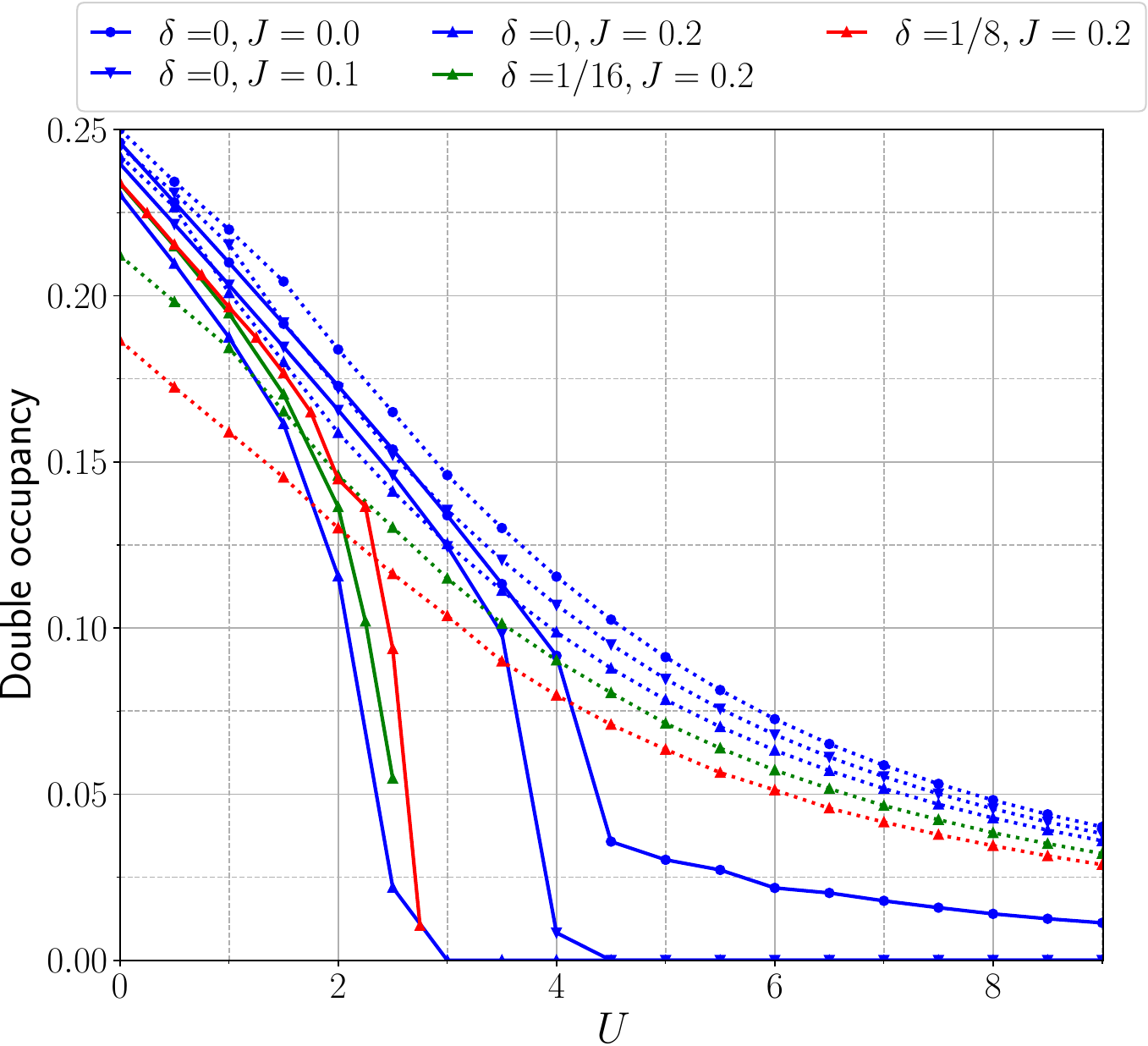}
      \caption{\textbf{Double occupancy vs $U$ for a $4\times 12$ cylindrical cluster.} The hole-doping values are $\delta=\{0,1/16,1/8\}$ and the spin exchange couplings are $J=\{0,0.1,0.2\}$. The solid (dotted) curves show the slave-spin-1 (direct DMRG) results.}
  \label{fig:Docc_tUVJ_vs_U_J_finite}
\end{figure}

\subsection{Double occupancy}

In Fig.~\ref{fig:Docc_tUVJ_vs_U_J_finite}, we first show the electron double occupancy $D$ averaged over all sites:
\begin{align}
\label{eq:double_occupancies}
D = \frac{1}{\mathcal{N}}\sum_i\langle\hat{n}_{i,\uparrow}\hat{n}_{i,\downarrow}\rangle,
\end{align}
\noindent
with the notation $\langle\cdot\rangle$ emphasizing that the expectation value is taken in the many-body electron ground state of Eq.~\eqref{eq:tUVJ_Ham}. In the slave-spin case, $D = \frac{1}{2\mathcal{N}}\sum_i\langle\hat{S}_{i}^z(\hat{S}_{i}^z-1)\rangle_s$. 

\begin{figure*}[htb!]
  \centering
    \includegraphics[width=\textwidth]{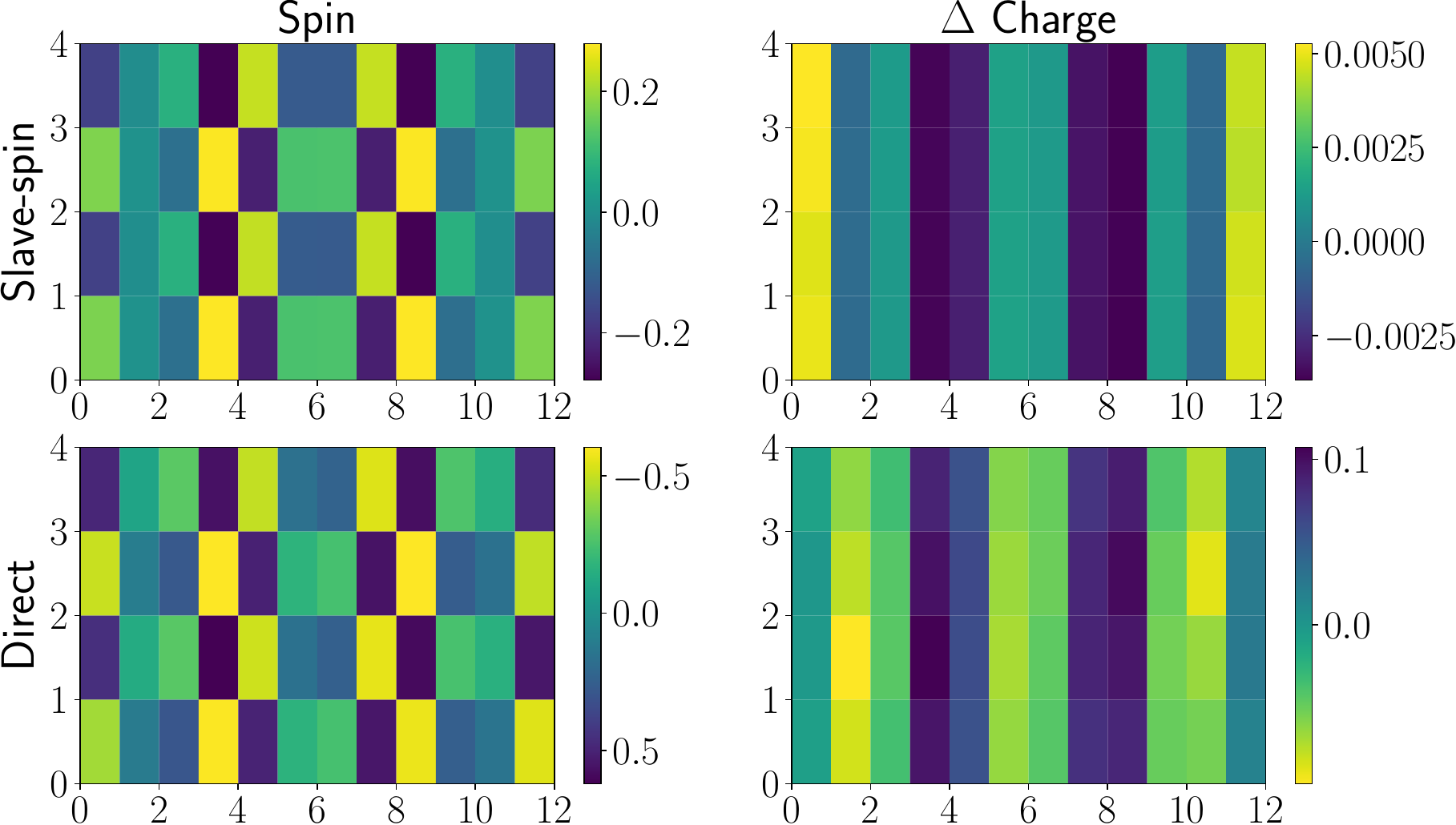}
      \caption{\textbf{Spin- and charge-density waves for $\delta=1/4$ and $V=0$.} In the panels on the left, the spin magnetization $M$ is shown, while in the right ones the electron density (pseudo-spin magnetic moment) about the average density per site is shown. The top panels show the slave-spin solution for $J=0.2$ and $U=2.5$. The bottom panels show the direct DMRG solution of the $t$-$U$-$V$-$J$ model for $J=0.4$ and $U=5.5$. The clusters of size $12\times4$ are periodic along the shortest side, \textit{i.e.}, $L_y$.}
  \label{fig:tot_docc_tUVJ_vs_U_45}
\end{figure*}

Focusing first on the slave-spin results, we observe very sharp drops in $D$ that occur in the neighborhood of $U_c$ for clusters of size $4\times12$, periodic along the longest side $L_y$. Overall, the curve profiles match those of $\Phi$ plotted in Fig.~\ref{fig:Phi_vs_J_U_zoom_J_0p1}, since both the double occupancy $D$ and $\Phi$ in the slave-spin-1 theory capture the local charge correlations of the original fermions. Like reported in Refs.~\cite{PhysRevLett.83.3498,PhysRevB.66.165111}, $D$ stays finite, but small, in the Mott-insulating regime at $J=0$ as $U\to\infty$. At $J\neq0$ and $\delta=0$, in the Mott-insulating regime, $D\to0$ passed $U_c$. $D$ decreases with increasing $J$ at fixed $U$ and fixed doping. This behavior can also be seen in the slave-spin-1 theory through the decrease of the crossover $U_c$, beyond which the system becomes poorly conducting or insulating, when $J$ is increased. This means that the growth of $J$ is related to an overall depletion of doublons, favoring nearest-neighbor sites to have opposite spins---two opposites spins on one site would, on top of costing an energy $\sim U$, be penalized by $J$. Note that adding a NN interaction $V$ would decrease further the double occupancies for all values of $U$ and $\delta$ (not shown). Indeed, if two opposite-spin electrons occupy a single site, they pay an extra energy amount $\sim V$ due to proximal electrons. At finite $J$, when doping away from $\delta=0$, we do not converge to solutions that enforce the physical constraints \eqref{eq:constraint_eqs_2} at larger values of $U$.

We turn our attention to the estimations of $D$ (Eq.~\eqref{eq:double_occupancies}) calculated by directly solving Eq.~\eqref{eq:tUVJ_Ham} with DMRG, by focusing on the dotted curves in Fig.~\ref{fig:Docc_tUVJ_vs_U_J_finite}. We observe that doping away from half-filling reduces $D$. This is in contrast with the slave-spin results, where at fixed $J=0.2$, when doping away from $\delta=0$, we find $D$ to slightly increase. Moreover, the very sharp drops in $D$ in the vicinity to the (doped) Mott crossover ($U_c$) are absent in the direct simulations. At half-filling, at fixed $U$, increasing the AFM coupling magnitude $J$ reduces the double occupancy $D$, just like what slave-spin theory predicts. 

Thanks to the cluster embedding, the slave-spin-1 method can access the thermodynamic limit. This constitutes one of the main differences between the slave-spin-1 results and those obtained from direct DMRG, and it allows for U(1) symmetry breaking through the field $\Phi$ coupled to the environment. As a consequence, the slave-spin approach exhibits sharp drops in $D$ at the crossover $U_c$. 

\subsection{Stripes}

In Fig.~\ref{fig:tot_docc_tUVJ_vs_U_45}, for $12\times4$ clusters, we illustrate both the magnetization per site, $M\equiv\langle\hat{n}_{i,\uparrow}-\hat{n}_{i,\downarrow}\rangle$, and the electron density per site, in the left and right panels, respectively. There, in the top and bottom panels, we show the results obtained by solving Eq.~\eqref{eq:tUVJ_Ham} at hole doping $\delta=1/4$ and $V=0$ by means of the slave-spin-1 theory outlined in Sec.~\ref{sec:self_consistent_cluster_approximation_main}, and by directly employing DMRG, respectively. In the top (bottom) panels of Fig.~\ref{fig:tot_docc_tUVJ_vs_U_45}, the on-site interaction $U=2.5$ ($U=5.5$) and the NN spin exchange $J=0.2$ ($J=0.4$). A direct mapping of the parameters used in Fig.~\ref{fig:tot_docc_tUVJ_vs_U_45} to study the physics within the slave-spin theory would locate $U$ and $J$ in the bad metal region of the slave-spin phase diagram. For the two methods used, we notice that a charge (resp. spin) density wave forms with period of $\sim2$ (resp. $\sim 4$) sites. In the left panels, the SDW is observed with the fading---associated to lower absolute magnetic moments---of the checkerboard spin background. In the right panels, blue color regions mean that more electrons aggregate there, while yellow-green regions show hole aggregations. The main difference between the two methods lies in the amplitudes of the stripes that are generally reduced in the slave-spin theory (top panels) when compared to the direct DMRG results (bottom panels). Thus, albeit on smaller cluster sizes, we do capture phased-in stripes in both the spin and charge sectors. The period of the stripes does decrease even more when the hole doping is increased from $\delta=1/16$ and $\delta=1/8$ to $\delta=1/4$ (see Figs.~\ref{fig:CDW_SDW} and \ref{fig:CDW_SDW_2}).

\begin{figure}[htb]
    \centering
    \includegraphics[width=\columnwidth]{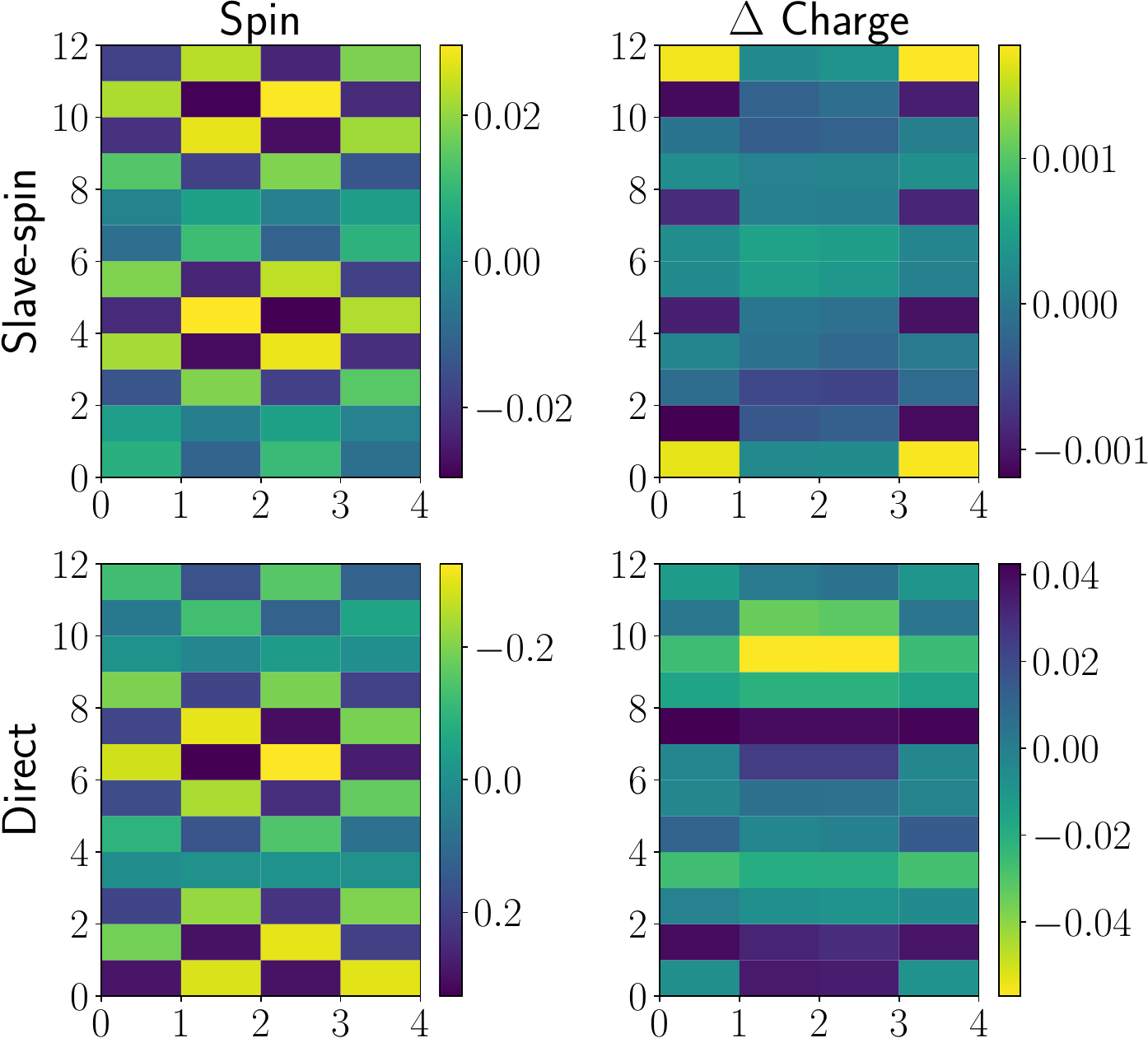}
    \caption{\textbf{SDWs and CDWs at $\delta=1/8$ for $4\times12$ cylindrical clusters}. The top two panels show results obtained from the slave-spin method, while the bottom ones show that of the direct DMRG simulations. The panels on the left (spin) illustrate the spin-1/2 patterns, and the panels on the right (charge) show the deviations of the charge density in real-space about its average. In the direct DMRG simulations (bottom panels), the color bar shows the electron filling oscillations about it average, with blue (yellow) meaning larger (lower) densities of electrons. The colors in the color bar associated to the slave-spin results (top panels) mean the same thing based off the mapping \eqref{eq:original_fermion_mappings}. In the top panels, $U=1.25$, $J=0.2$ and $V=0$, and in the bottom panels, $U=2.5$, $J=0.2$ and $V=0$.}
    \label{fig:rotated_clusters_direct_DMRG}
\end{figure}

We then look at some rotated version of the cylinder cluster considered in Fig.~\ref{fig:tot_docc_tUVJ_vs_U_45}, in order to establish if the slave-spin can adequately capture some key features captured by DMRG.
We now deal with a cylindrical cluster size $4\times12$ (periodic along the longest side $L_y$). In the top panels of Fig.~\ref{fig:rotated_clusters_direct_DMRG}, we illustrate how the spin (left panel) and charge (right panel) densities distribute into space within the slave-spin-1 framework for $U=1.25$, $J=0.2$, $V=0$ and $\delta=1/8$. In the bottom panels, using the same panel layout, $V$, $J$ and $\delta$, we show direct DMRG result of Eq.~\eqref{eq:tUVJ_Ham} for $U=2.5$. As far as the spin channel is concerned, everything looks the same, just shifted along the $y$-axis. Similar to Fig.~\ref{fig:tot_docc_tUVJ_vs_U_45}, the amplitudes of the stripes are reduced in the slave-spin formalism compared to the direct DMRG calculations. In the charge sector, the patterns are also shifted between the two cases. They however differ more than in the spin sector. Nevertheless, in both cases, nascent CDWs seem to emerge; we have seen that increasing the system size along $L_y$ and/or $L_x$ favor the stabilization of stripes.

We have seen in Sec.~\ref{sec:self_consistent_mf_solution} that the stripes and doped-insulator physics are qualitatively captured by the slave-spin-1 cluster mean-field approach developed in Sec.~\ref{sec:Methods}.
As shown in Appendix~\ref{sec:cluster_geometry_size}, the key features of the stripes show up regardless of the cluster size and geometry, as long as only one side is periodic to induce symmetry breaking in the pseudo-spin sector. The first kinetic term of Eq.~\eqref{eq:4_sites_square_impurity} needs to be mean-field-decoupled along the open sides to estimate $\Phi$. Open boundary conditions of the cluster do not stabilize CDWs nor SDWs. Cylinders that are periodic along the longest side are geometries that diminish significantly the computational cost of the slave-spin simulations, allowing to scale up to larger system sizes with fewer resources. 


\section{Conclusion}
\label{sec:conclusion}

In this work, we used a slave-spin-1 formalism to treat the $t$-$U$-$V$-$J$ model for fermions. In this formalism, the original fermion field is decoupled into a spin-1/2 pseudo-fermion treated by means of inhomogeneous mean-field theory, representing the spin degrees of freedom, and a pseudo-spin-1 XXZ transverse-field system, representing the charge degrees of freedom. Through the evaluation of 2- and 4-point correlation functions in the respective sectors and their self-consistent exchanges across the sectors that renormalize the microscopic couplings, interesting physics emerged, such as stripes and (doped) Mott physics. 

We studied the buildup of charge- and spin-density waves upon hole doping. In the weak $U$ limit, for values of $J\geq0.15$, coherent (phased-in) commensurate charge- and spin-density waves appear (filled stripes). The periodicity of those stripes increase with the reduction of the hole doping. At higher dopings, like $\delta=1/8$, the stripe wavelength is more-or-less equal across $J$ and $U$. This formalism captures qualitatively the growth of stripe wavelength with $1/\delta$; we see that doubling the doping halves the wavelength. Adding a nearest-neighbor interaction $V$ enhances the amplitude of the charge stripes, as reported in Ref.~\cite{Abram_2017}.

The slave-spin-1 theory also gives rise to (doped) Mott-insulating phases, where the impact of magnetic and charge correlations, coming from $J$, $U$ and $V$, was studied. We showed that the onset of the doped Mott insulator shifts to higher values of $U$ when doping in holes. The opposite phenomenon is observed when $V$ or $J$ is cranked up. On the one hand, $V$ contributes in the localization of the pseudo-spins, thereby diminishing the electron quasiparticle weight. On the other hand, $J$ renormalizes down the in-plane spin coupling term of quantum XXZ model, thereby facilitating the out-of-plane ordering of the pseudo-spins, which coincides with the development of the bad metal phase.

The formalism introduced here for the $t$-$U$-$V$-$J$ could be used to develop hybrid quantum-classical algorithms, for example, synthetic cold atom quantum simulators~\cite{PhysRevB.109.174409,2025hybridquantumclassicalanalogsimulation}, and open up the opportunity to study other exotic, yet still misunderstood physics, exhibited in the high-T$_c$ cuprates~\cite{Bednorz1986,RevModPhys.92.031001,RevModPhys.82.2421}, transition-metal dichalcogenides 1T-TaSe$_2$~\cite{doi:10.1073/pnas.1706769114,PhysRevResearch.2.013099,PhysRevX.7.041054} or organic charge-transfer crystals $\kappa$-(BEDT-TTF)$_2$~\cite{PhysRevLett.91.107001,Yamashita2008,Yamashita2009}.
Indeed, through the usage of this slave-particle decoupling of the fermion field degrees of freedom~\cite{PhysRevB.66.165111,PhysRevB.70.035114}, neutral-atom platforms implementing anisotropic quantum spin models, like the quantum XY~\cite{Chen2023,PhysRevLett.132.263601,Bornet2023}, XXZ~\cite{PRXQuantum.3.020303}, and XYZ~\cite{PhysRevLett.130.243001} models, could be used to map fermionic Hamiltonians such as the $t$-$U$-$V$-$J$ model for various electron fillings and lattice geometries. The complex quantum spin model could be implemented onto a Rydberg quantum simulator~\cite{Labuhn2016,PRXQuantum.3.020303}, using qutrits~\cite{PRXQuantum.6.020332}. While our focus in this work remains theoretical, recent advances in coherent control~\cite{f58h-zxs3} raise the possibility that effective slave-particle descriptions of correlated fermion models could eventually be implemented and tested in such quantum simulators.

Future steps include the study of the nonequilibrium dynamics and the effects of temperature of such a theory via the usage of time-dependent minimally entangled typical thermal states (METTS)~\cite{wang2025spectroscopycomplextimecorrelations,PhysRevLett.102.190601}.
One could also look at the nonthermal dynamics of the doped Mott insulator, at charge and spin stripes in a photodoped scenario~\cite{Murakami2022}, and at the effect of adding a next-nearest-neighbor hopping amplitude. Finally, one could think of treating the neglected U(1) gauge fluctuations stemming from the slave-spin decoupling. This should be possible in DMRG~\cite{PhysRevD.98.074503,Bañuls2013} and Rydberg cold atom platforms~\cite{Cheng2024,PRXQuantum.3.040317,PhysRevX.4.031027}.

\begin{acknowledgments}
We thank Fabien Alet, Antoine Browaeys, Luca de' Medici, Serge Florens, Antoine Georges and Thierry Lahaye for stimulating and insightful discussions. O.S. acknowledges support from the Swiss National Science Foundation through the Postdoc.Mobility fellowship. This work was granted access to the HPC resources of TGCC and IDRIS under the allocation A0190510609 attributed by GENCI (Grand Equipement National de Calcul Intensif). It has also used high performance computing resources of IDCS (Infrastructure, Données, Calcul Scientifique) under the allocation CPHT 2025. This work is part of HQI initiative (\url{www.hqi.fr}) and is supported by France 2030 under the French National Research Agency award number ANR-22-PNCQ-0002 and ANR-22-EXES-0013.
\end{acknowledgments}

\bibliographystyle{apsrev4-1_our_style}
\bibliography{BibDoc}

\clearpage


\appendix

\renewcommand\thefigure{S.\arabic{figure}}  \setcounter{figure}{0} 

\section{Constraint validity}
\label{sec:constraint_validity}

Like detailed in Sec.~\ref{sec:tJUV_model}, we need to introduce Lagrange multipliers to confine the solutions within the physical Hilbert space for the slave-spin formalism when calculating observable averages.

In this section, we check out the validity of the approximation embodied in the local constraints \eqref{eq:constraint_eqs_2}, when it comes to constraining the action of the slave-spin field operators to the physical Hilbert space. In relation to the locality of the Lagrangian multiplier, we carry out the demonstration for a single-site Hubbard model, narrowing down to the following set of equations per sector~\cite{PhysRevB.66.165111} ($V=J=0$):

\begin{subequations}
  \begin{align}
    \begin{aligned}
    \label{eq:decoupled_equations_imp_Ham_f} 
      \hat{H}^f_{U} &= \epsilon_0\sum_{\sigma}\hat{f}_\sigma^\dagger\hat{f}_\sigma - \mu_f\sum_{i,\sigma}\hat{n}^f_{i,\sigma}
    \end{aligned}\\
    \begin{aligned}
     \label{eq:decoupled_equations_imp_Ham_theta}
      \hat{H}^{s}_{U} = \frac{U}{2}\sum_i\hat{S}^{z}_{i}\left(\hat{S}^{z}_{i}-1\right) - \mu_s\sum_{i}\hat{S}^z_{i}.
    \end{aligned}
  \end{align}
\end{subequations}
\noindent
The set of equations~\eqref{eq:decoupled_equations_imp_Ham_f} and \eqref{eq:decoupled_equations_imp_Ham_theta} can be easily deduced from those describing the $t$-$U$-$V$-$J$ model Eqs.~\eqref{eq:decoupled_equations_Ham_f} and \eqref{eq:decoupled_equations_Ham_theta}. As for the kinetic term, since we are considering a single site, the tight-binding hopping term in Eq.~\eqref{eq:decoupled_equations_Ham_f} is traded for a single energy level $\epsilon_0$. In this particular model, the pseudo-fermion sector deals specifically with the kinetic energy term while the pseudo-spin sector deals with the potential energy term.

\begin{figure}[htb]
  \centering
    \includegraphics[width=\columnwidth]{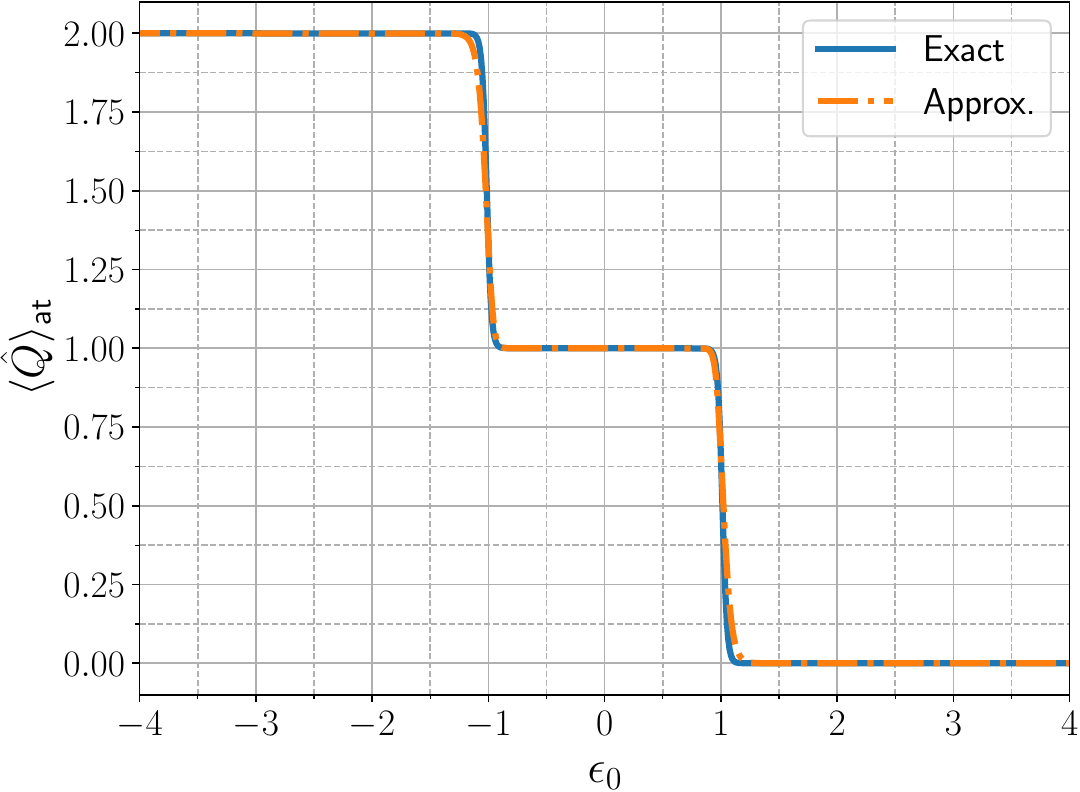}
      \caption{\textbf{Coulomb staircase of the spin-1 slave-particle theory} applied to the SU(2) impurity Hubbard model. $\langle Q\rangle_{at}$ denotes the average charge density on the impurity site, while $\epsilon_0$ represents the local impurity energy. An inverse temperature $\beta=50$ is used, at $U=2$.}
  \label{fig:Coulomb_staircase_N_2}
\end{figure}

We aim to verify that in the atomic limit, the Coulomb staircase, describing the original electron occupation number $\langle \hat{Q}\rangle_{\text{at}}\equiv \sum_{\sigma\in\{\uparrow,\downarrow\}}\langle\hat{c}^{\dagger}_{\sigma}\hat{c}_{\sigma}\rangle_{\text{at}}$ as a function of the local impurity energy, is recovered. The expectation value $\langle\cdot\rangle_{\text{at}}$ signifies that operators operate on eigenstates of the single-site impurity Hubbard model, with the physical electron ladder operators $\hat{c}^{(\dagger)}$ defined on the impurity. To obey the constraints \eqref{eq:constraint_eqs_2}, we require that $\langle\hat{S}^z_i\rangle_s = \langle\hat{n}_i^f\rangle_f-1$ be satisfied on average. In the atomic limit, this translates to 

\begin{align}
\label{eq:constraint_imp_Hubbard_1}
&\langle\hat{S}^z_i\rangle_s = \langle\hat{n}_i^f\rangle_f-1\notag\\
&\implies\frac{1}{\mathcal{Z}_s}\sum_{m=-1}^1me^{-\beta E_m} = 2n_F(\epsilon_0-\mu_f)-1,
\end{align}
\noindent
where $m$ corresponds to the eigenvalues of the spin-1 $\hat{S}^z$ operator, $\beta$ the inverse temperature, $E_m = \frac{U}{2}m^2+\left(\mu_s-\frac{U}{2}\right)m$, and $\mathcal{Z}_s=\sum_{m=-1}^1e^{-\beta E_m}$ is the partition function of the pseudo-spin sector. In Eq.~\eqref{eq:constraint_imp_Hubbard_1}, the Fermi-Dirac distribution function is denoted $n_F$, and $\mu_f$ and $\mu_s$ play the role of the Lagrange multipliers. In Fig.~\ref{fig:Coulomb_staircase_N_2}, as an orange dash-dotted line, we show that the transcendental equation \eqref{eq:constraint_imp_Hubbard_1} matches very well the exact average occupation number $\langle \hat{Q}\rangle_{\text{at}}$ computed from the impurity Hubbard model (in blue). Indeed, for all values of $\epsilon_0$, there exists a simultaneous solution to $\mu_f$ and $\mu_s$ which reproduces faithfully the strong local charge correlations (Coulomb staircase). The exact value of the Coulomb staircase reads

\begin{align}
\label{eq:atomic_limit_Z_chargon}
\langle \hat{Q}\rangle_{\text{at}} = \frac{1}{\mathcal{Z}_\text{at}}\sum_{Q=0}^2\begin{pmatrix}2\\Q\end{pmatrix}Qe^{-\beta E_Q}, 
\end{align}
with $E_Q=\epsilon_0Q + \frac{U}{2}Q\left(Q-1\right)$ corresponding to the $3$ eigenvalues of the SU(2) single-site impurity Hubbard model, endowed with $\begin{pmatrix}2\\Q\end{pmatrix}$ degeneracies. The partition function $\mathcal{Z}_\mathrm{at}$ is simply 

\begin{align*}
\mathcal{Z}_\text{at}=\sum_{Q=0}^2\begin{pmatrix}2\\Q\end{pmatrix}e^{-\beta E_Q}.
\end{align*}

In Fig.~\ref{fig:Coulomb_staircase_N_2}, the blue solid curve shows Eq.~\eqref{eq:atomic_limit_Z_chargon} as a function of orbital-site energy $\epsilon_0$.

\section{Pseudo-fermion inhomogeneous mean field}
\label{sec:inhomogeneous_MF}

In this section, the inhomogeneous self-consistent mean-field equations of the spin super-exchange term in Eq.~\eqref{eq:decoupled_equations_Ham_f} are laid out and the procedure is explained, following closely the procedures employed in Ref.~\cite{PhysRevB.108.035139}. In particular, we want to determine more precisely the content of Eq.~\eqref{eq:simplified_pseudo_fermion_term}. The interacting Hamiltonian term in the pseudo-fermion sector reads (second term of Eq.~\eqref{eq:decoupled_equations_Ham_f})

\begin{align}
\label{eq:tJ_inhomogeneous_MF}
\hat{H}^{f}_{tUVJ} &= \sum_{\langle ij\rangle} \tilde{J}_{ij} \ \mathbf{S}^f_i\cdot\mathbf{S}^f_j\notag\\ 
&= \sum_{\langle ij\rangle} \tilde{J}_{ij} \ \hat{f}^{\dagger}_{i,\alpha}\boldsymbol{\tau}_{\alpha\beta}\hat{f}_{i,\beta}\cdot\hat{f}^{\dagger}_{j,\gamma}\boldsymbol{\tau}_{\gamma\delta}\hat{f}_{j,\delta}\notag\\
&= \sum_{\langle ij\rangle, \sigma} \tilde{J}_{ij} \biggl[\bigl(\hat{n}^{f}_{i,\sigma}\hat{n}^{f}_{j,\sigma}-\hat{n}^{f}_{i,-\sigma}\hat{n}^{f}_{j,\sigma}\bigr)\notag\\
&\phantom{========}+ 2\hat{f}^{\dagger}_{i,\sigma}\hat{f}_{i,-\sigma}\hat{f}^{\dagger}_{j,-\sigma}\hat{f}_{j,\sigma}\biggr].
\end{align}
\noindent
We defined in Eq.~\eqref{eq:tJ_inhomogeneous_MF} $\tilde{J}_{ij}\equiv J\langle\hat{S}_i^-\hat{S}^+_i\hat{S}_j^-\hat{S}^+_j\rangle_s$, thereby absorbing the 4-point correlation function into the spin exchange coupling. The vector of Pauli matrices is again defined as $\boldsymbol{\tau}$. If we assume that the fluctuations in particle density are small compared to some mean field, meaning that $\hat{n}^f_{i,\sigma}\to \langle \hat{n}^f_{i,\sigma}\rangle_f + \delta \hat{n}_{i,\sigma}$, with $\delta \hat{n}_{i,\sigma}\equiv \hat{n}^f_{i,\sigma} - \langle \hat{n}^f_{i,\sigma}\rangle_f$, one obtains

\begin{align}
\label{eq:tJ_inhomogeneous_MF_1}
&\hat{H}^{f,\text{MF},(1)}_{tUVJ} \notag\\
&=\sum_{\langle ij\rangle, \sigma} \tilde{J}_{ij} \biggl[\underbrace{\langle\hat{n}^{f}_{i,\sigma}-\hat{n}^{f}_{i,-\sigma}\rangle_f}_{\equiv M_i^{\sigma}}\hat{n}^{f}_{j,\sigma}+\left(\hat{n}^{f}_{i,\sigma}-\hat{n}^{f}_{i,-\sigma}\right)\langle\hat{n}^{f}_{j,\sigma}\rangle_f\notag\\
&\phantom{==} - \langle\hat{n}^{f}_{i,\sigma}\rangle_f\langle\hat{n}^{f}_{j,\sigma}\rangle_f + \langle\hat{n}^{f}_{i,-\sigma}\rangle_f\langle\hat{n}^{f}_{j,\sigma}\rangle_f\biggr].
\end{align}
\noindent
In Eq.~\eqref{eq:tJ_inhomogeneous_MF_1}, we dropped out the terms featuring prefactors $\delta \hat{n}^2$, since they are negligible owing to the approximation. We also defined the magnetization $M_i^{\sigma}$. 

We can carry on with other significant fluctuation channels that can arise in the Hamiltonian \eqref{eq:tJ_inhomogeneous_MF}. For that, let us rewrite the Hamiltonian \eqref{eq:tJ_inhomogeneous_MF}:

\begin{align}
\label{eq:tJ_inhomogeneous_MF_rewritten}
\hat{H}^{f}_{tUVJ} &= \sum_{\langle ij\rangle, \sigma} \tilde{J}_{ij} \biggl[\hat{f}^{\dagger}_{i,\sigma}\hat{f}_{j,-\sigma}\hat{f}^{\dagger}_{j,-\sigma}\hat{f}_{i,\sigma}-\hat{f}^{\dagger}_{i,\sigma}\hat{f}_{j,\sigma}\hat{f}^{\dagger}_{j,\sigma}\hat{f}_{i,\sigma}\notag\\
&\phantom{========}+ 2\hat{f}^{\dagger}_{i,\sigma}\hat{f}_{i,-\sigma}\hat{f}^{\dagger}_{j,-\sigma}\hat{f}_{j,\sigma}\biggr].
\end{align}
\noindent
We then decouple in Eq.~\eqref{eq:tJ_inhomogeneous_MF_rewritten} the mean field parameters $\hat{\Delta}_{i,j}^{+,\sigma}\equiv -\hat{f}^{\dagger}_{i,\sigma}\hat{f}_{j,-\sigma} = (\hat{\Delta}_{j,i}^{-,\sigma})^{\dagger}$, as well as $\hat{\Delta}_{i,j}^{z,\sigma}\equiv -\hat{f}^{\dagger}_{i,\sigma}\hat{f}_{j,\sigma} = (\hat{\Delta}_{i,j}^{z,\sigma})^{\dagger}$, just like it was done for the density operator (without a hat, it means that this is the expectation value). We start dealing with terms leading to nonzero expectation values of $\hat{\Delta}^z$ (terms that do not induce spin flip upon hopping). The set of terms figuring in Eq.~\eqref{eq:tJ_inhomogeneous_MF_rewritten} that comply with this requirement are

\begin{align}
\label{eq:tJ_inhomogeneous_MF_rewritten_delta_z}
\hat{H}^{f}_{tUVJ}[\hat{\Delta}^{z}] &= -\sum_{\langle ij\rangle, \sigma} \tilde{J}_{ij} \biggl[\hat{f}^{\dagger}_{i,\sigma}\hat{f}_{j,\sigma}\hat{f}^{\dagger}_{j,\sigma}\hat{f}_{i,\sigma}\notag\\
&\phantom{========}+ 2\hat{f}^{\dagger}_{i,\sigma}\hat{f}_{j,\sigma}\hat{f}^{\dagger}_{j,-\sigma}\hat{f}_{i,-\sigma}\biggr],
\end{align}
\noindent
which yields, after performing the mean-field decoupling on $\hat{\Delta}^z$'s:

\begin{align}
\label{eq:tJ_inhomogeneous_MF_2}
&\hat{H}^{f,\text{MF},(2)}_{tUVJ} = \sum_{\langle ij\rangle, \sigma} \tilde{J}_{ij} \biggl[\Delta_{ij}^{z,\sigma}\left(\hat{f}_{i,\sigma}^{\dagger}\hat{f}_{j,\sigma}+\text{H.c}\right)\notag\\
&+2\Delta_{ij}^{z,\sigma}\left(\hat{f}_{i,-\sigma}^{\dagger}\hat{f}_{j,-\sigma} + \text{H.c}\right)+\left(\Delta^{z,\sigma}_{ij}\right)^2+2\Delta_{ij}^{z,\sigma}\Delta_{ji}^{z,-\sigma}\biggr].
\end{align}

We then set out to deal with terms leading to nonzero expectation values of $\hat{\Delta}^{\pm}$ (terms that induce spin flip upon hopping). This requirement means we perform the mean-field decoupling on these terms

\begin{align}
\label{eq:tJ_inhomogeneous_MF_rewritten_delta_pm}
\hat{H}^{f}_{tUVJ}[\hat{\Delta}^{\pm}] &= \sum_{\langle ij\rangle, \sigma} \tilde{J}_{ij} \biggl[\hat{f}^{\dagger}_{i,\sigma}\hat{f}_{j,-\sigma}\hat{f}^{\dagger}_{j,-\sigma}\hat{f}_{i,\sigma}\notag\\
&\phantom{========}+ 2\hat{f}^{\dagger}_{i,\sigma}\hat{f}_{i,-\sigma}\hat{f}^{\dagger}_{j,-\sigma}\hat{f}_{j,\sigma}\biggr],
\end{align}
\noindent
yielding:

\begin{align}
\label{eq:tJ_inhomogeneous_MF_3}
&\hat{H}^{f,\text{MF},(3)}_{tUVJ} = -\sum_{\langle ij\rangle, \sigma} \tilde{J}_{ij} \biggl[\Delta_{ij}^{+,\sigma}\hat{f}_{j,-\sigma}^{\dagger}\hat{f}_{i,\sigma} + \hat{f}^{\dagger}_{i,\sigma}\hat{f}_{j,-\sigma}\Delta_{ji}^{-,\sigma}\notag\\
&\phantom{====}+ 2\Delta_{ii}^{+,\sigma}\hat{f}^{\dagger}_{j,-\sigma}\hat{f}_{j,\sigma} + 2\hat{f}_{i,\sigma}^{\dagger}\hat{f}_{i,-\sigma}\Delta_{jj}^{-,\sigma}\notag\\
&\phantom{====}+ \Delta_{ij}^{+,\sigma}\Delta_{ji}^{-,\sigma} + 2\Delta_{ii}^{+,\sigma}\Delta_{jj}^{-,\sigma} \biggr].
\end{align}
\noindent
The total mean-field Hamiltonian, which sums up Eq.~\eqref{eq:simplified_pseudo_fermion_term}, is the combination of thrice Eqs.~\eqref{eq:tJ_inhomogeneous_MF_1}, \eqref{eq:tJ_inhomogeneous_MF_2} and \eqref{eq:tJ_inhomogeneous_MF_3}; $\hat{H}^{f,\text{MF}}_{tUVJ} = \sum_k \hat{H}^{f,\text{MF},(k)}_{tUVJ}$. If on-site spin-flip is forbidden, the terms proportional to $\hat{\Delta}^{\pm}_{ii}$ zero out when acting on the ground state. $\hat{H}^{f,\text{MF},(3)}_{tUVJ}$ does not play a significant role.

The self-consistent gap equations are the following:

\begin{align}
\label{eq:self_consistent_gap_eq}
\begin{cases}
M_{i}^{\sigma} &= \langle\hat{n}^{f}_{i,\sigma}-\hat{n}^{f}_{i,-\sigma}\rangle_f\\
\Delta_{ij}^{\pm,\sigma} &= -\langle\hat{f}^{\dagger}_{i,\sigma}\hat{f}_{j,-\sigma}\rangle_f\\
\Delta_{ij}^{z,\sigma} & = -\langle\hat{f}^{\dagger}_{i,\sigma}\hat{f}_{j,\sigma}\rangle_f.
\end{cases}
\end{align}
\noindent
We note that $\Delta_{ij}^{+,\uparrow}=\Delta_{ji}^{-,\downarrow}$---the conjugate of this equation also holds. These mean-field parameters in Eq.~\eqref{eq:self_consistent_gap_eq} are updated from $2L_x\times L_y\equiv 2\mathcal{N}$ eigenvectors $v_{i,\sigma}^l$, $l\in[1,\cdots,2\mathcal{N}]$, diagonalizing $\langle\hat{H}^{f,\text{MF}}_{tUVJ}\rangle_f$ of size $2\mathcal{N}\times2\mathcal{N}$:

\begin{align}
\label{eq:update_MF_equations}
\langle\hat{f}^{\dagger}_{i,\sigma}\hat{f}_{j,\sigma^{\prime}}\rangle_f = \sum_{l=1}^{2\mathcal{N}} (v_{i,\sigma}^l)^{\ast}v_{j,\sigma^{\prime}}^ln_F(\epsilon_l-\mu_f),
\end{align}
\noindent
where $n_{F}$ is the Fermi-Dirac distribution function, and $\mu_f$ the chemical potential. At finite temperature, the total density of pseudo-fermions can be calculated from 

\begin{align}
\label{eq:update_MF_equations}
n = \frac{1}{\mathcal{N}}\sum_i^{\mathcal{N}}\langle\hat{n}^{f}_{i}\rangle_f = \frac{1}{\mathcal{N}}\sum_{l=1}^{2\mathcal{N}}n_F(\epsilon_l-\mu_f),
\end{align}
\noindent
where $\epsilon_l$ are the $2\mathcal{N}$ eigenvalues. Within this inhomogeneous mean-field ansatz, based off the self-consistent gap equations \eqref{eq:self_consistent_gap_eq}, and the three mean-field components Eqs.~\eqref{eq:tJ_inhomogeneous_MF_1}, \eqref{eq:tJ_inhomogeneous_MF_2} and \eqref{eq:tJ_inhomogeneous_MF_3}, the expectation value $\langle \mathbf{S}^f_i\cdot\mathbf{S}^f_j\rangle_f$ yields

\begin{align}
\label{eq:expectation_val_SiSj}
\langle \mathbf{S}^f_i\cdot\mathbf{S}^f_j\rangle_f &= \sum_{\sigma}\langle \hat{n}_{i,\sigma}\rangle_f\left(\langle \hat{n}_{j,\sigma}\rangle_f - \langle \hat{n}_{j,-\sigma}\rangle_f\right)\notag\\
& - \sum_{\sigma}\biggl(\Delta_{ij}^{z,\sigma}\Delta_{ji}^{z,\sigma} + 2\Delta_{ij}^{z,\sigma}\Delta_{ji}^{z,-\sigma}-\Delta_{ij}^{+,\sigma}\Delta_{ji}^{-,\sigma}\notag\\
&-2\Delta_{ii}^{+,\sigma}\Delta_{jj}^{-,\sigma}\biggr).
\end{align}

\section{Slave-spin-1 theory for homogeneous mean fields}
\label{app:Hubbard_slave_rotor}

In this Appendix, we work out the self-consistent slave-rotor mean-field method discussed in Sec.~\ref{sec:self_consistent_cluster_approximation_main} for the special case of homogeneous mean fields. In the same philosophy as detailed in Ref.~\cite{PhysRevB.76.195101}, we wish to solve for the ground state of a Hamiltonian on a finite plaquette coupled to an external mean-field. That would be done by diagonalizing the Hamiltonian on a plaquette and coupling it self-consistently to the rest of the lattice. Remarkably, according to Ref.~\cite{PhysRevB.76.195101}, incorporating the short-range correlations on the plaquette would dramatically improve the predictability of the self-consistent mean-field theory; for instance, in the triangular lattice Hubbard model a spin liquid is observed at half-filling on a finite cluster and not on an impurity. It is essential to keep in mind that the slave-rotors, used in Ref.~\cite{PhysRevB.76.195101} and truncated down to angular momenta $l=1$, are objects defining an infinite-dimensional algebra. Here, we do not make any such approximations and use spin-1 objects to perform the spin-charge decoupling.
 
\subsection{Pseudo-spin sector}
\label{sec:rotor_sector}

For the following example, let us suppose that we are dealing with a $2\times2$ cluster size embedded self-consistently within a mean-field environment, as shown in Fig.~\ref{fig:4_x_4_lattice}. Even though we are dealing with larger lattices, the examples carries over to larger sizes and various cluster geometries straightforwardly. The local basis states for the four-site cluster can be expressed as $\ket{S_1^{z},S_2^{z},S_3^{z},S_4^{z}}_s$, with the local Hamiltonian corresponding to Eq.~\eqref{eq:decoupled_equations_Ham_theta} with slight modifications to parameters representing the mean-field environment where the NN $\chi_{ij}$ in Eq.~\eqref{eq:decoupled_equations_Ham_theta} is defined as $\chi$, as well as NN electron hopping $t_{ij}$ defined as $t$. 

Like discussed in the main text, the Mott transition is characterized by $\Phi=0$ and a charge gap at half-filling, whereas a nonzero $\Phi$ means that charge can hop in and out of the cluster and hence the charge can fluctuate (correlated metal). The homogeneous mean-field treatment of the pseudo-fermion sector entails a loss in real-space resolution of the correlation functions passed from one sector to the other (see Fig.~\ref{fig:flow_chart_scheme}). Upon convergence of the parameters in the pseudo-spin sector, the $B_{ij}$ correlator expectation values are computed within the pseudo-spin ground state and averaged over the NN bonds:

\begin{align}
\label{eq:B_nearest_neighbor_correlator}
B_{\langle i,j\rangle} \equiv B = \overline{\langle \hat{S}^+_i\hat{S}^-_j\rangle}_s,
\end{align}
where the bar means that the correlators are averaged. The 4-point correlation function $\langle\hat{S}_i^-\hat{S}^+_i\hat{S}_j^-\hat{S}^+_j\rangle_s$ is also averaged over the NN bonds of the $2\times2$ cluster. The correlators are then passed on to the pseudo-fermion sector once $\mu_{s}$, and $\Phi$ have converged in the ground state $\ket{\Psi_s}$.

\begin{figure}[htb]
  \centering
    \includegraphics[width=\columnwidth]{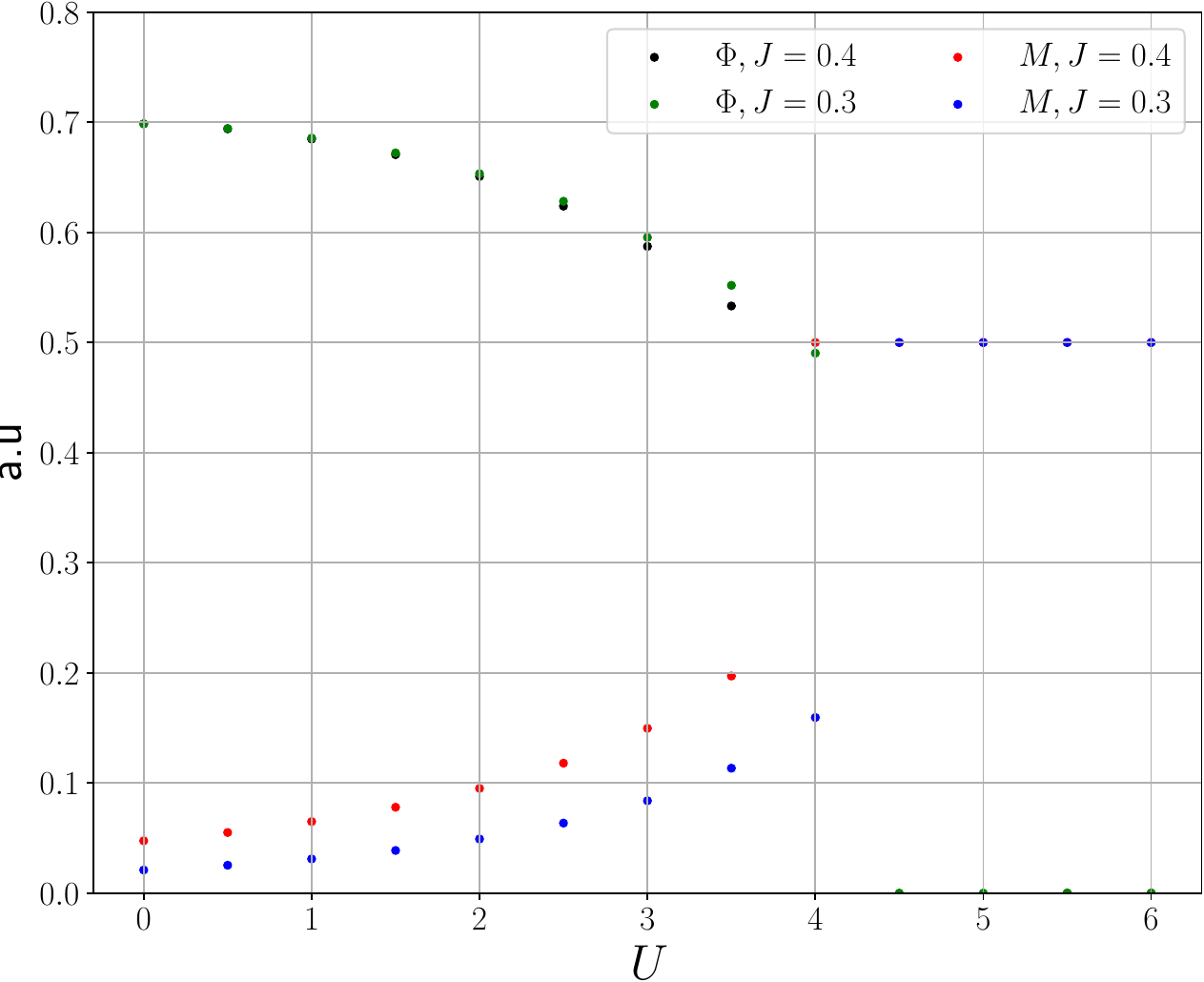}
      \caption{\textbf{Metal-insulator order parameter $\Phi$ and on-site magnetization $M$ as a function of $U$} for a small $2\times2$ cluster using homogeneous self-consistent mean-field slave-spin-1 theory. The values of $J=0.3$ and $0.4$.}
  \label{fig:Homogeneous_MF_slave_spin_1}
\end{figure}

\subsection{Pseudo-fermion sector}
\label{sec:spinon_sector}

Then, we look back at the pseudo-fermion Hamiltonian \eqref{eq:decoupled_equations_Ham_f}. Here, we solve for $\chi$ (NN) to inject back those into the pseudo-spin self-consistency problem explained in Sec.~\ref{sec:rotor_sector}. Note that one could make use of higher-order neighboring connections---beyond NN, if the cluster permits. Fourier transforming Eq.~\eqref{eq:decoupled_equations_Ham_f}, one gets

\begin{align}
\label{eq:k_space_spinon_hamiltonian}
\hat{H}_f = \sum_{\mathbf{k},\sigma}\left(\epsilon_{\mathbf{k}} - \mu_f\right)\hat{f}_{\mathbf{k},\sigma}^{\dagger}\hat{f}_{\mathbf{k},\sigma} + J\tilde{S}\overline{\langle\hat{S}_i^-\hat{S}^+_i\hat{S}_j^-\hat{S}^+_j\rangle}_s,
\end{align}
where the renormalized pseudo-fermion dispersion relation for the square lattice reads

\begin{align}
\label{eq:spinon_dispersion_relation}
&\epsilon_{\mathbf{k}} = -2tB\left(\cos{k_x} + \cos{k_y}\right),
\end{align}
where the lattice spacing $a\equiv1$. The extra term in Eq.~\eqref{eq:k_space_spinon_hamiltonian} comes from the mean-field decoupling of the term $\propto\hat{\mathbf{S}}^f_i\cdot\hat{\mathbf{S}}^f_j$ in the second term of Eq.~\eqref{eq:decoupled_equations_Ham_f}:

\begin{align}
\label{eq:AFM_decoupling_homo_MF}
\hat{\mathbf{S}}^f_i\cdot\hat{\mathbf{S}}^f_j &\to -\frac34\chi_{ij}\left(\hat{f}^{\dagger}_{i,\sigma}\hat{f}_{j,\sigma}+\text{H.c}\right) \notag\\
&+ (-1)^{i+j}\frac{M}{2}\left(\hat{n}^f_{j,\uparrow}-\hat{n}^f_{j,\downarrow}-\hat{n}^f_{i,\uparrow}+\hat{n}^f_{i,\downarrow}\right)\notag\\
&\simeq -\frac32\chi^2-M^2\notag\\
&\equiv\tilde{S},
\end{align}
\noindent
where $M\equiv\langle\hat{n}^f_{\uparrow}-\hat{n}^f_{\downarrow}\rangle_f$ is the magnetization, homogeneous throughout the lattice. The chemical potential $\mu_f$ is tuned so as to match the number of electrons (see last line of Eq.~\eqref{eq:constraint_eqs_2}):

\begin{align}
\label{eq:electron_doping}
1-\delta = \frac{1}{N}\sum_{\mathbf{k}}\Theta\left(-\epsilon_{\mathbf{k}}+\mu_f\right),
\end{align}
since we are at zero temperature. The Heaviside function is denoted by $\Theta$. 

We now elaborate on the definition of the correlation function $\chi_{i,j}$ introduced in Eq.~\eqref{eq:decoupled_equations_Ham_theta}. Given that pseudo-fermions obey fermionic statistics, we can define $\chi_{i,j}$ as

\begin{align}
\label{eq:spinon_statistics_chi_ij}
\chi_{i,j} = \lim_{\tau\to 0^-}\frac{1}{\beta}\sum_{n=-\infty}^{\infty} e^{-i\omega_n\tau} \frac{1}{\mathcal{N}}\sum_{\mathbf{k}} e^{i\mathbf{k}\cdot\left(\mathbf{r}_i-\mathbf{r}_j\right)}\mathcal{G}_{\mathbf{k}}^f(i\omega_n),
\end{align}
where $0<\tau<\beta$ is the imaginary time, with $\beta$ the inverse temperature, $\omega_n\equiv (2n+1)\pi/\beta$ are the Matsubara frequencies, $\mathcal{N}$ is the number of sites, and $\mathcal{G}_{\mathbf{k}}^f$ is the noninteracting pseudo-fermion propagator defined within the original Brillouin zone (BZ). At zero temperature, using both the components of the renormalized pseudo-fermion dispersion relation \eqref{eq:spinon_dispersion_relation} and Eq.~\eqref{eq:spinon_statistics_chi_ij}, the variable $\chi$ can be updated accordingly

\begin{align}
\label{eq:chi_chi_prime_update}
\chi &= -\frac{1}{\mathcal{N}}\sum_{\mathbf{k}}\Theta\bigl(-\epsilon_{\mathbf{k}}+\mu_f\bigr)\frac{\epsilon_{\mathbf{k}}}{2tB}.
\end{align}
The chemical potential $\mu_f$ appearing in Eq.~\eqref{eq:chi_chi_prime_update} is the one that converged to the sought electron density, determined via Eq.~\eqref{eq:electron_doping} using some root-finding method.

\begin{figure}[htb]
  \centering
    \includegraphics[scale=0.45]{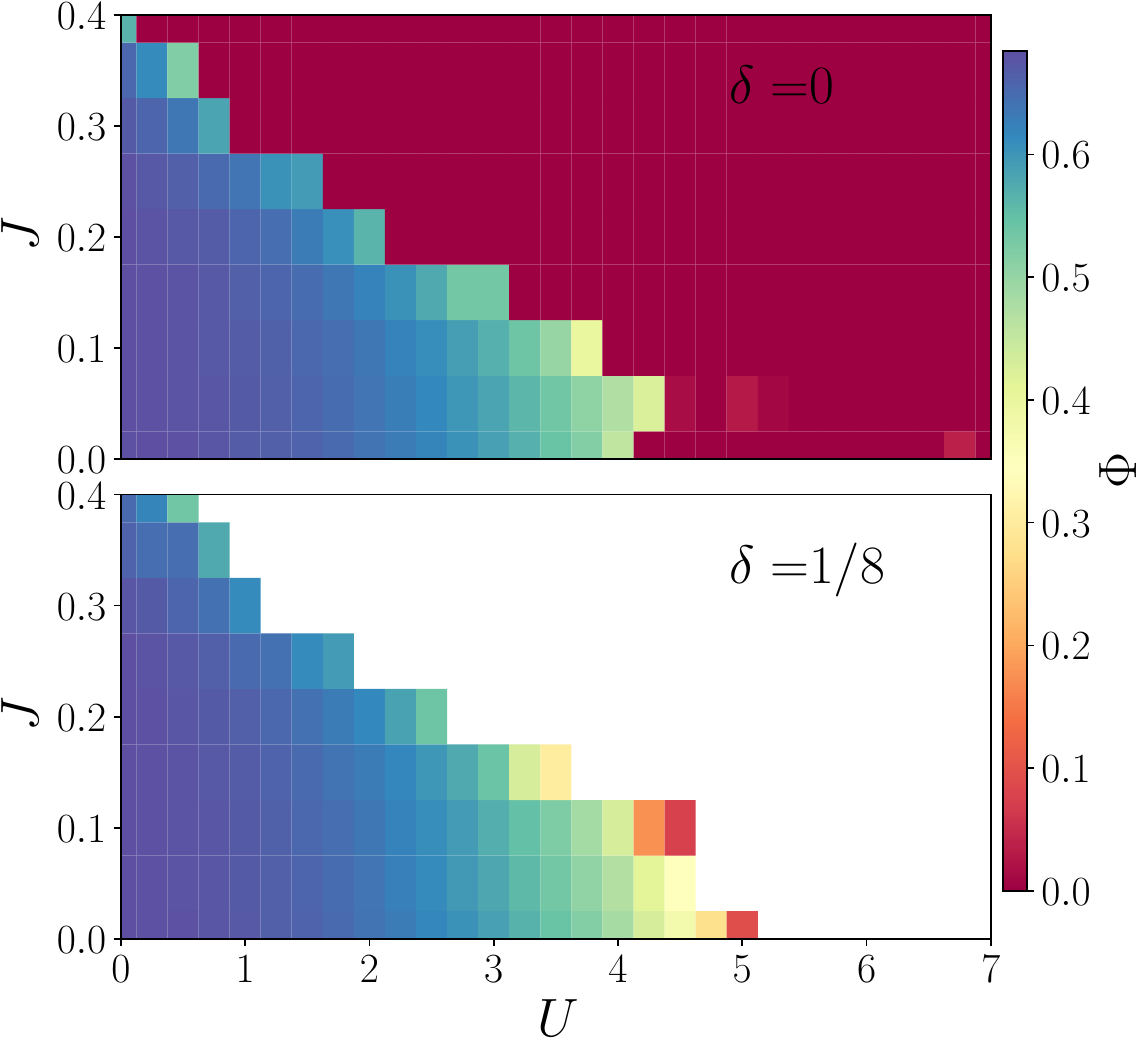}
      \caption{\textbf{Metallic order parameter $\Phi$ vs $J$ and $U$ for various hole dopings.} The value of $V=0$. Top panel: $\delta=0$. Bottom panel: $\delta=1/8$. The color bar shows the intensity of $\Phi$. The calculations were carried out on a $4\times 16$ lattice with periodic boundary conditions along the longest side $L_y$. Again, the white region means that the slave-spin solution has not converged.}
  \label{fig:Phi_vs_J_U_4_32}
\end{figure}

Afterwards, once the chemical potential $\mu_f$ and $\chi$ have been calculated, these are fed back to the pseudo-spin sector. The whole scheme, depicted in Fig.~\ref{fig:flow_chart_scheme}, is iterated until overall convergence. In Fig.~\ref{fig:Homogeneous_MF_slave_spin_1}, we show the result of the phase diagram for the slave-spin-1 using homogeneous mean field theory for the square lattice $t$-$U$-$J$ model. We show the magnetization $M$ and metal-insulator parameter $\Phi$ as a function of $U$. We observe that once $M\to1/2$, $\Phi\to0$ for all $J$'s considered. The filing here corresponds to half-filling ($\delta=0$).

\section{Cluster geometry and size}
\label{sec:cluster_geometry_size}

In this section of the Appendix, we show some results accompanying those shown in the main text for cylindrical cluster sizes $4\times16$ and $4\times31$, $6\times12$ and $8\times8$. The goal here is to show that we do observe stripes whatever the clusters considered and the conclusions in the main text do not hold only for long narrow cylinders. (The clusters must be periodic on one side).

We first check the map of $\Phi$ in the $J-U$ parameter space for a cluster size $4\times 16$ periodic along $L_y$. This can be contrasted directly with Fig.~\ref{fig:Phi_vs_J_U}, bottom panels. Thus, in Fig.~\ref{fig:Phi_vs_J_U_4_32}, we show in the top panel the metal-insulator parameter $\Phi$ at half-filling, while in the bottom panel we show $\Phi$ at hole-doping $\delta=1/8$.

Similarly to the cluster size $4\times32$, we reach the same conclusions pertaining to the growth of local charge correlations and NN spin correlations; as $J$ increases, the onset of the (doped) Mott-insulating regime is shifted down to lower values of $U$. Moreover, doping in holes away from half-filling extends out in $U$ the metallic regime.

\begin{figure*}[htb]
  \centering
    \includegraphics[width=\textwidth]{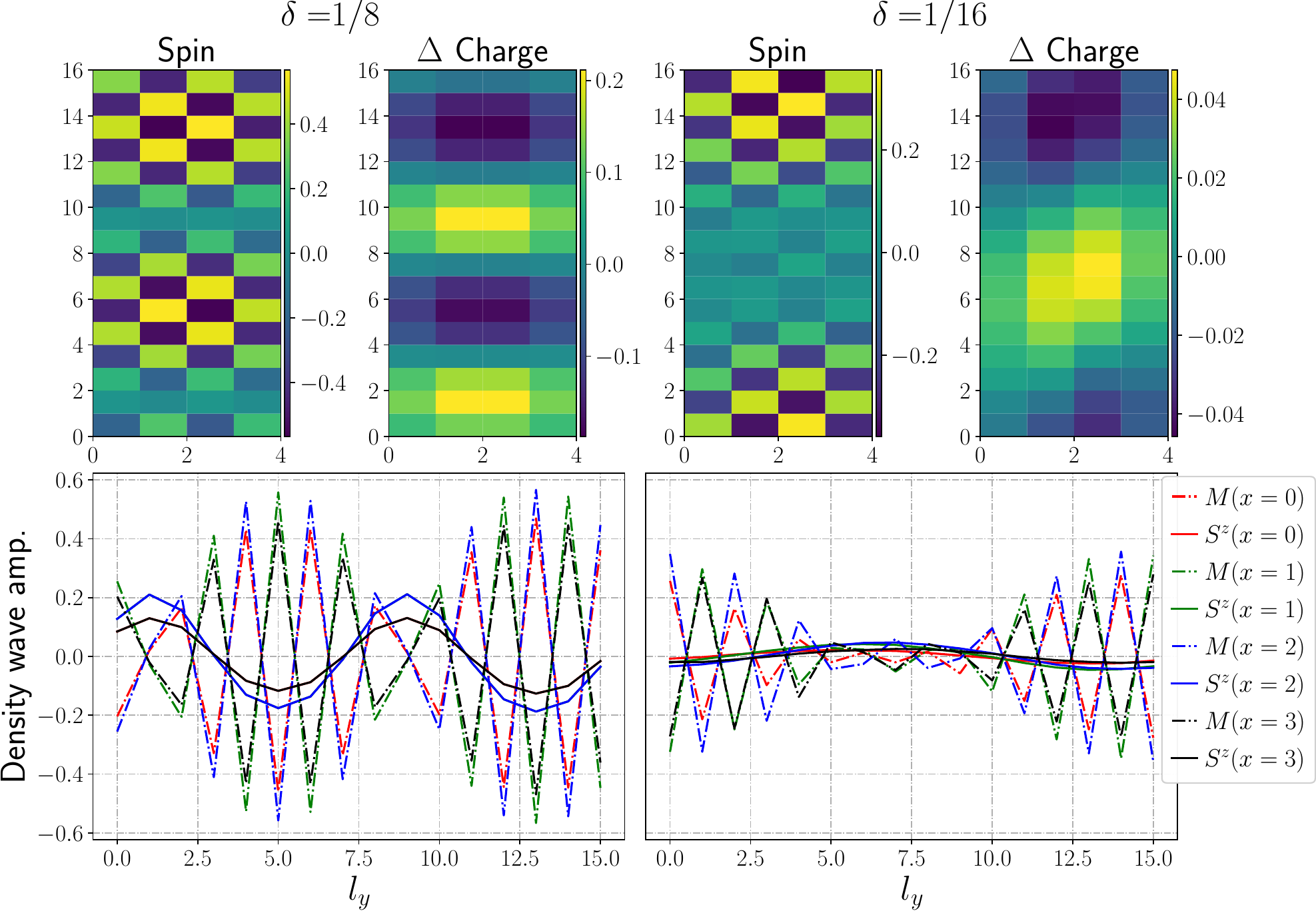}
      \caption{\textbf{Real-space signatures of the SDWs and CDWs as a function of doping at $J=0.2$ and $V=0$.} Upper left panel: magnetization showing staggered AFM ordering in the pseudo-fermion sector (left inset) and pseudo-spin-1 modulations about the average showing CDWs (right inset) --- the yellow (blue) color relates to surplus of holes (electrons) with respect to half-filling. The value of $\delta=1/16$ and $U=1.5$. Upper right panel: same as upper left panel, although for values of $\delta=1/8$ and $U=2$. Bottom left panel: pseudo-fermion magnetization $M$ along strips $l_x$ of 16 sites (dotted curves) and spin-1 projections $S^z$ (solid curves) for the same set of parameters as the panel above ($\delta=1/16$ and $U=1.5$). Bottom right panel: same data presented as that of the bottom left panel, although for parameters $\delta=1/8$ and $U=2$. For readability, the charge oscillations are amplified by a factor $10$ for both panels. This figure can be directly compared to Fig.~\ref{fig:CDW_SDW} for a larger system size.}
  \label{fig:CDW_SDW_2}
\end{figure*}

In Fig.~\ref{fig:CDW_SDW_2}, we make a direct comparison with Fig.~\ref{fig:CDW_SDW} for cylindrical clusters of size $4\times16$. The rest of the parameters are identical, namely the values of $U$, $V$ and $\delta$. Hence, the left set of three panels (top and bottom ones) shows $\delta=1/8$, $U=2$ and $J=0.2$, while the set of three panels on the right display results for $\delta=1/16$, $U=1.5$ and $J=0.2$. Even with longer lattice rings, we do clearly capture the decrease of the period of the SDWs and CDWs upon doping away from half-filling. On average, for CDWs, we observe a period of $\sim7-8$ sites for $\delta=1/8$ while we see a period of $\sim13-14$ sites for $\delta=1/16$---the periods are double for SDWs. Thus, upon doubling the presence of holes, the periodicity is halved. We can see that the finite-size effect is more noticeable at doping $\delta=1/16$, where the pseudo-spin modulation at $l_x=0,1$ shifts with respect to that at $l_x=2,3$.

\begin{figure}[h!]
  \centering
    \includegraphics[scale=0.3]{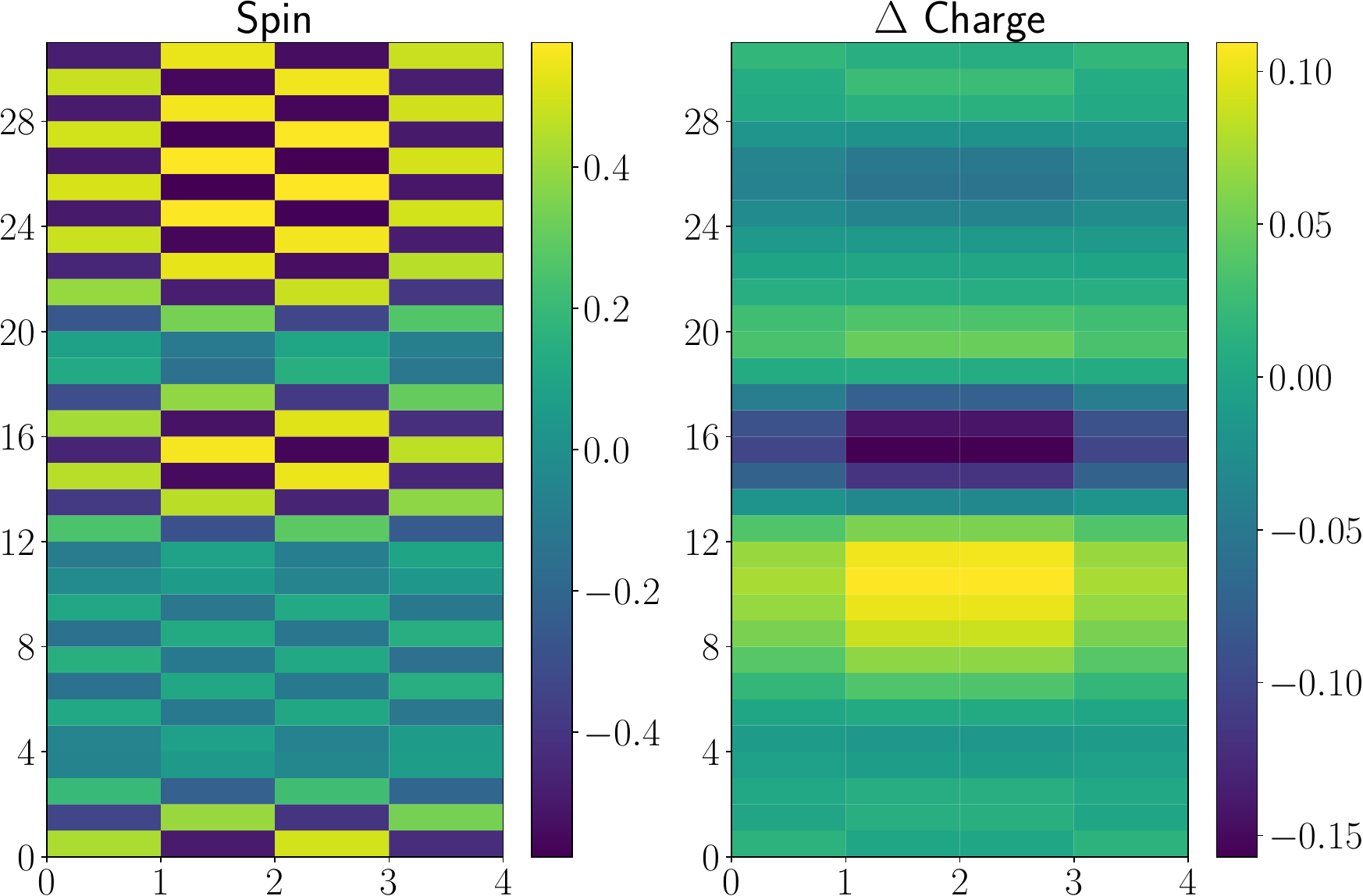}
      \caption{\textbf{SDWs and CDWs at incommensurate doping $\delta=1/8$}. The NN interaction $V=0$, $J=0.2$ and $U=2$. The system size is $4\times31$. Left panel: Magnetization as a function of the lattice sites. Right panel: deviations from average of the pseudo-spin-1 magnetic moment for all cluster sites.}
  \label{fig:CDWs_SDWs_incommensurate}
\end{figure}

In the following, we show that it is important to ensure that the doping be commensurate with the system size. In Fig.~\ref{fig:CDWs_SDWs_incommensurate}, we show results for a cylindrical cluster $4\times31$ at $\delta=1/8$. In this case, the number of sites is not divisible by $8$, therefore the equivalent of electrons removed from half-filling is not an integer value. We show results for $U=2$, $J=0.2$ and $V=0$, just like in Figs.~\ref{fig:CDW_SDW} and \ref{fig:CDW_SDW_2}. We see in Fig.~\ref{fig:CDWs_SDWs_incommensurate} that the charge and spin stripes are offset from each other, and that the period is affected when compared to cluster sizes commensurate with the filling. As long as the filling is commensurate with the amount of sites, the stripes are phased in.

\begin{figure*}[htb]
    \centering
    \begin{subfigure}{0.47\textwidth}
        \includegraphics[scale=0.3]{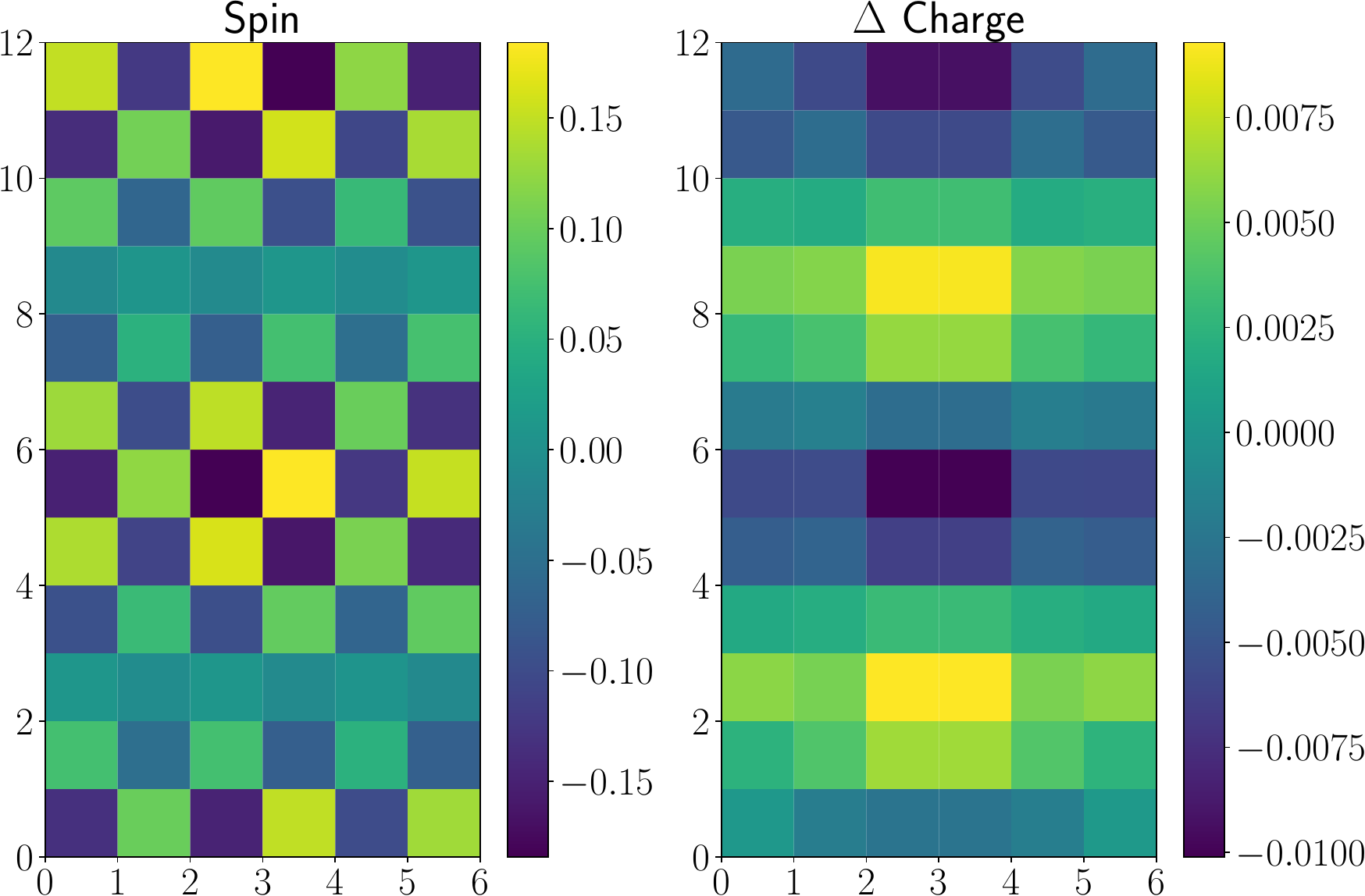}
        \caption{\textbf{SDWs and CDWs at $\delta=1/4$ for a cylindrical $6\times12$ cluster}. $V=0$, $J=0.2$ and $U=2$.}
    \end{subfigure}
    \quad\quad
    \begin{subfigure}{0.47\textwidth}
        \includegraphics[scale=0.3]{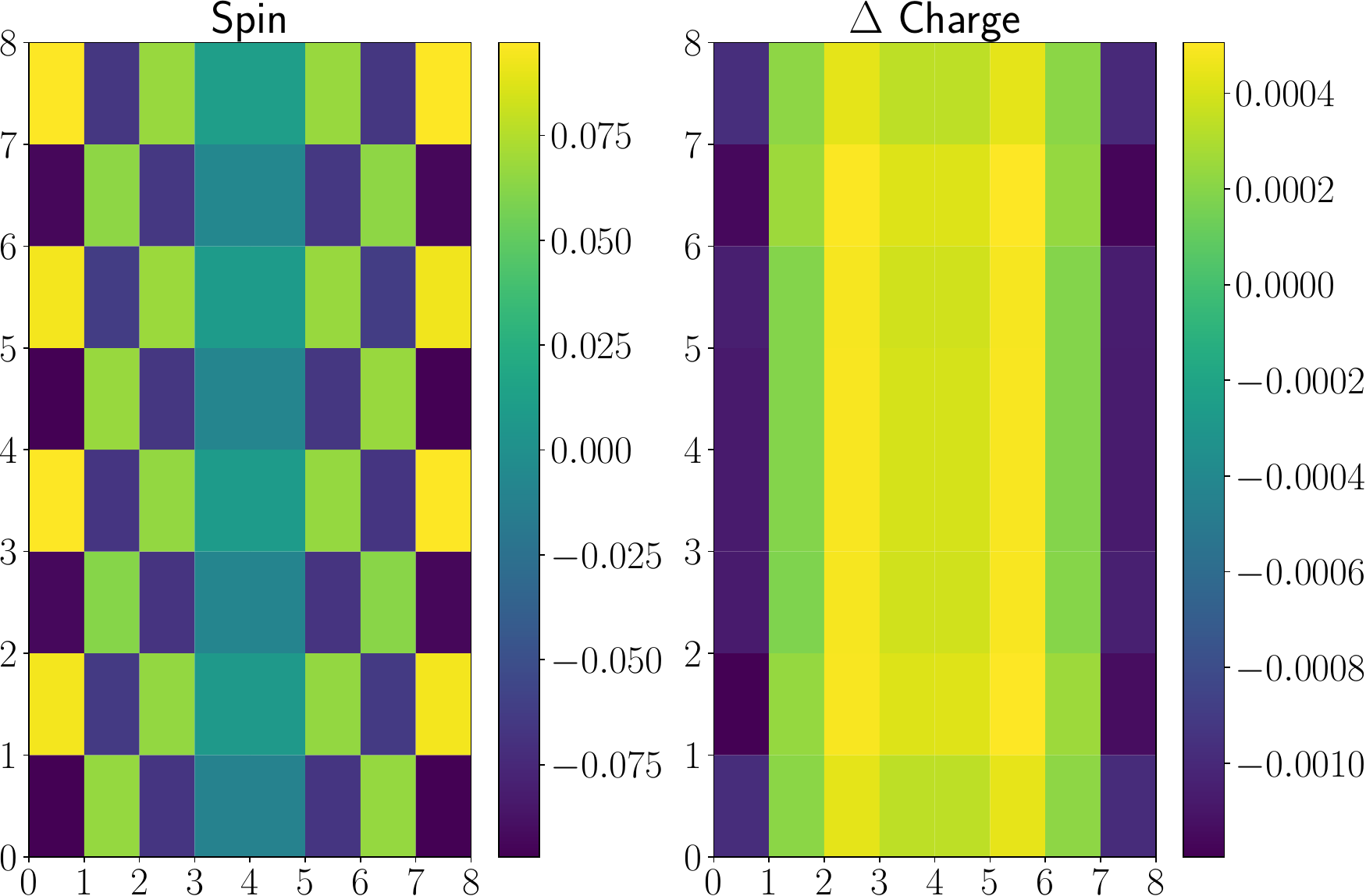}
        \caption{\textbf{SDWs and CDWs at $\delta=1/8$ for a cylindrical $8\times8$ cluster}. $V=0$, $J=0.1$ and $U=3$.}
    \end{subfigure}
    \caption{\textbf{SDWs and CDWs for various cluster sizes and dopings}. For each subfigure, the left panel shows the real-space magnetization, while the right panel shows the variation about the average of the pseudo-spin magnetic moment related to the charge filling.}
    \label{fig:app_various_cluster_sizes}
\end{figure*}

In Fig.~\ref{fig:app_various_cluster_sizes}, we show different sizes of cluster so as to demonstrate that stripes are not only seen on narrow clusters when using the slave-spin-1 approximation. We emphasize that the usage of long narrow cylinders was motivated by the lower computational cost, allowing to map out complete phase diagrams at many dopings. In the left subplot of Fig.~\ref{fig:app_various_cluster_sizes}, we show the magnetization (left panel) and pseudo-spin quantum number (right panel) for a $6\times12$ cylindrical cluster at $\delta=1/4$, $U=2$ and $J=0.2$. The stripes obtained here are quite similar, although a bit shorter in period, to those seen in Fig.~\ref{fig:CDW_SDW_2}, left sub-figure, where $\delta=1/8$, $U=2$ and $J=2$ for a $4\times16$ cylindrical cluster. Then in the right sub-figure of Fig.~\ref{fig:app_various_cluster_sizes}, we show the stripe textures for $8\times8$ cluster at $\delta=1/8$, $U=3$ and $J=0.1$. In this case, the stripes rotate by $\frac\pi2$ to span out horizontally just like for regular cylindrical clusters (see Fig.~\ref{fig:tot_docc_tUVJ_vs_U_45}).

\end{document}